\documentclass[journal=jacsat,manuscript=communication,etalmode=truncate,keywords=true]{achemso}
\setkeys{acs}{articletitle=true,maxauthors=10,etalmode=truncate,keywords=true}
\usepackage{physics}
\usepackage{mathptmx}
\usepackage{amsmath,amssymb}
\usepackage{amsfonts}
\usepackage{dcolumn}
\usepackage{graphicx,MnSymbol}
\usepackage{helvet,wasysym}
\usepackage[bookmarks=true,colorlinks=true,urlcolor=black,linkcolor=black,citecolor=blue]{hyperref} 
\usepackage[usenames,dvipsnames]{color}
\usepackage{colortbl,cuted}
\usepackage{balance}

\definecolor{JPCCBlue}{RGB}{34,80,169}
\definecolor{NLRed}{RGB}{215,24,30}
\definecolor{abstractcolor}{RGB}{255,243,201}
\definecolor{myblue}{RGB}{232,232,232}
\makeatletter\newenvironment{abstractbox}{%
   \begin{lrbox}{\@tempboxa}\begin{minipage}{0.988\textwidth}}{\end{minipage}\end{lrbox}%
   \colorbox{abstractcolor}{\usebox{\@tempboxa}}
}\makeatother

\usepackage{titlesec}
\titleformat{\section}{\bfseries\sffamily\color{JPCCBlue}}{\thesection.~}{0pt}{}
\titleformat{\subsection}[runin]{\bfseries\sffamily\normalsize}{\indent\thesubsection.~}{0pt}{}[.]
\titlespacing{\subsection}{0pt}{0pt}{*1}
\titleformat{\subsubsection}{\bfseries\sffamily\normalsize}{\thethesubsection.~}{0pt}{}
\titlespacing{\subsubsection}{0pt}{0pt}{*0}

\title{A hybrid quantum-classical algorithm for multichannel quantum scattering of atoms and molecules}

\author{Xiaodong Xing$^{1}$, Alejandro Gomez Cadavid$^{2,3,4}$, Artur F. Izmaylov$^{2,3}$, and Timur V. Tscherbul$^{1}$}

\affiliation{$^{1}$Department of Physics, University of Nevada, Reno, NV, 89557, USA\\
$^{2}$Chemical Physics Theory Group, Department of Chemistry, University of Toronto, Toronto, Ontario, M5S 3H6, Canada\\
$^{3}$Department of Physical and Environmental Sciences, University of Toronto Scarborough, Toronto, Ontario M1C 1A4, Canada\\
$^{4}$Kipu Quantum, Greifswalderstrasse 226, 10405 Berlin, Germany}

\email{ttscherbul@unr.edu} 

\begin{document}
\maketitle

\begin{strip}
\vspace{-1.cm}

\noindent{\color{JPCCBlue}{\rule{\textwidth}{0.5pt}}}
\begin{abstractbox}
\begin{tabular*}{17cm}{b{10.5cm}r}
\noindent\textbf{\color{JPCCBlue}{ABSTRACT:}}
We propose a hybrid quantum-classical algorithm for solving the time-independent Schr\"odinger equation for atomic and molecular  collisions. The algorithm is based on the $S$-matrix version of the Kohn variational principle, which computes the fundamental scattering $S$-matrix by inverting the Hamiltonian matrix expressed in the basis of square-integrable functions. The computational bottleneck of the classical algorithm---symmetric matrix inversion---is  addressed here using the variational quantum linear solver (VQLS), a recently developed noisy intermediate-scale quantum (NISQ)  algorithm for solving systems of linear equations. We apply our algorithm to single and multichannel quantum scattering problems, obtaining accurate vibrational relaxation probabilities in collinear atom-molecule collisions. We also show how the algorithm could be scaled up to simulate  collisions of large polyatomic molecules. Our results demonstrate that it is possible to calculate scattering cross sections and rates for complex molecular collisions on NISQ quantum processors, opening up the possibility of scalable digital quantum computation of gas-phase bimolecular collisions and reactions of relevance to  astrochemistry and ultracold chemistry.
&\includegraphics[scale=0.5,  trim = 18 0 0 0]{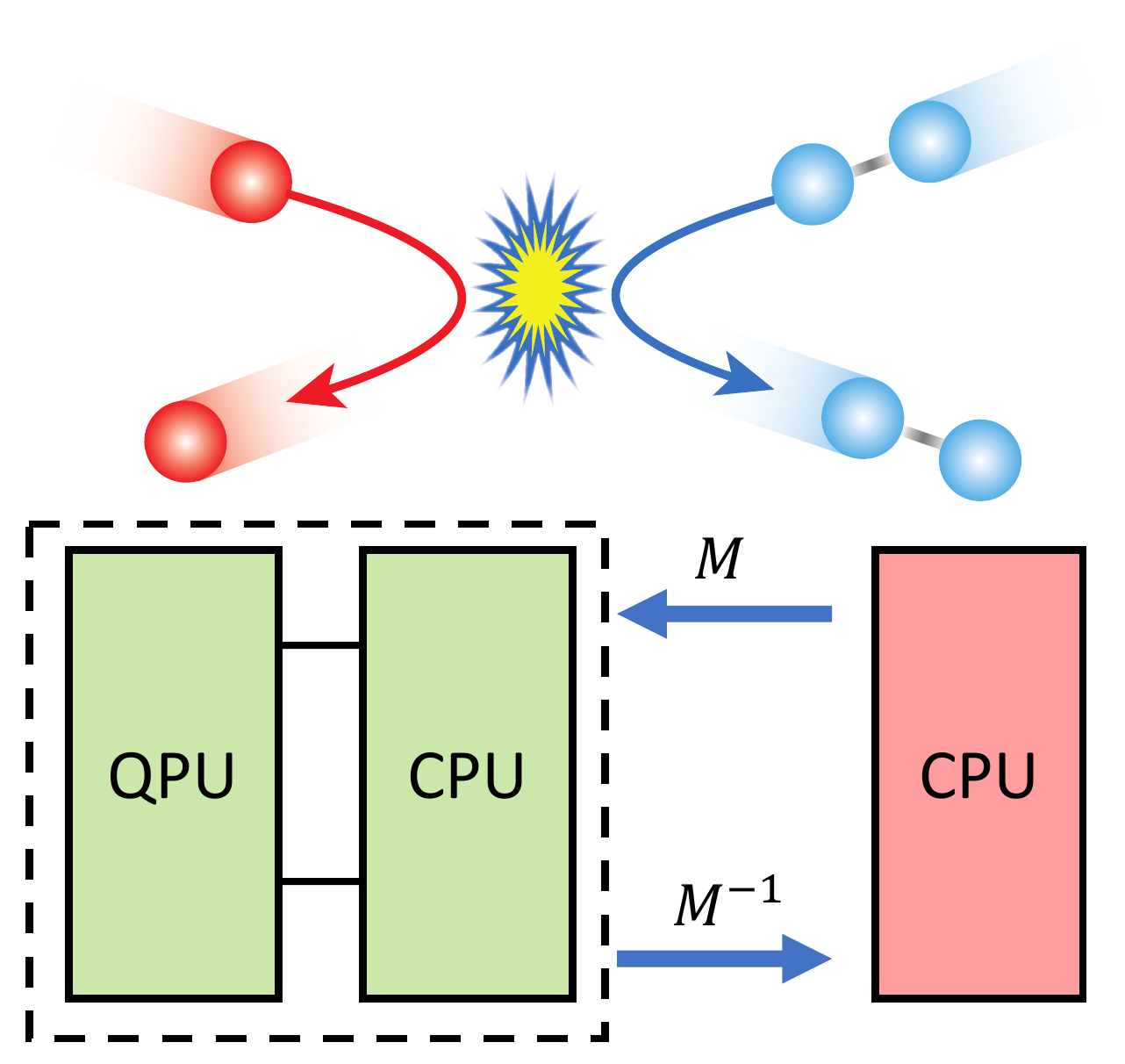}. \\
\end{tabular*}
\end{abstractbox}
\noindent{\color{JPCCBlue}{\rule{\textwidth}{0.5pt}}}
\end{strip}

\def\bigfirstletter#1#2{{\noindent
    \setbox0\hbox{{\color{JPCCBlue}{\Huge #1}}}\setbox1\hbox{#2}\setbox2\hbox{(}%
    \count0=\ht0\advance\count0 by\dp0\count1\baselineskip
    \advance\count0 by-\ht1\advance\count0 by\ht2
    \dimen1=.5ex\advance\count0 by\dimen1\divide\count0 by\count1
    \advance\count0 by1\dimen0\wd0
    \advance\dimen0 by.25em\dimen1=\ht0\advance\dimen1 by-\ht1
    \global\hangindent\dimen0\global\hangafter-\count0
    \hskip-\dimen0\setbox0\hbox to\dimen0{\raise-\dimen1\box0\hss}%
    \dp0=0in\ht0=0in\box0}#2}

The development of novel algorithms for the digital \cite{Cao:19,McArdle:20} and analog \cite{ArguelloLuengo:19,McArdle:20} quantum simulation of molecular structure and dynamics is a rapidly expanding field of research 
spanning a wide array of topics, including electronic structure  \cite{McArdle:20,Otten:2022,fan:2021,Colless:2018}, vibrational structure and spectroscopy \cite{McArdle:20,Otten:2022,fan:2021,Colless:2018}, and excitonic energy transfer in model biological systems \cite{Parrish:19,Leontica:21}.
Quantum phase estimation (QPE) \cite{QPE} is the foremost algorithm for solving the Schr{\"o}dinger equation on  fault-tolerant quantum computers  \cite{Nielsen:10,pezze:21}. While QPE can provide an exponential speedup over classical algorithms, current  hardware limitations  have prevented its widespread use. 
More promising for current applications are hybrid quantum algorithms, such as the variational quantum eigensolver (VQE)  \cite{Peruzzo:14,McClean:16,Romero:18}, which  combines classical and quantum computation to take full advantage of today's limited noisy intermediate-scale quantum (NISQ) resources \cite{Preskill:18}.  


The VQE algorithm consists of four steps \cite{Peruzzo:14,McClean:16,Romero:18} and begins with the preparation of a trial quantum state (or ansatz) $\ket{\psi({\theta_i})}$, followed by the calculation of the expectation value $\langle \hat{O}\rangle (\theta_i) =\bra{\psi(\theta_i)}\hat{O}\ket{\psi(\theta_i)}$ of a quantum observable $\hat{O}$ over the trial state. The optimal values of the variational parameters $\theta_i$ are then found using a classical optimization algorithm, and the entire procedure is repeated until convergence of the expectation value. 
The VQE has been successfully applied to the electronic structure problem \cite{Cao:19,McArdle:20} and to the calculation of molecular vibrational levels \cite{McArdle:19}, dynamics \cite{sparrow:18}, and spectra \cite{Sawaya:19}. 
An alternate approach  is based on mapping the vibrational eigenvalue problem  onto a quadratic unconstrained binary optimization problem, which could be solved efficiently on quantum annealers \cite{Johnson:11,Teplukhin:19}.
The resulting  quantum annealer eigensolver was used to obtain the vibrational energy levels of ozone  \cite{Teplukhin:19} and excited electronic states of NH$_3$ \cite{Teplukhin:21}.  Efficient quantum circuits have  been proposed for VQE  calculations of rovibrational energy levels in a discrete variable representation  basis \cite{Asnaashari:23}.

Quantum scattering phenomena play a fundamental role in physical and theoretical chemistry, being at the heart of gas-phase chemical reaction mechanisms \cite{Althorpe:03,Clary:08,Zhang:16}, astrochemistry \cite{Kaiser:02,Herbst:13}, combustion simulations \cite{Jasper:14,Klippenstein:17},
mode-selective chemistry \cite{Flynn:96,Liu:16} and atmospheric chemistry \cite{Yang:19c}.
In addition, molecular scattering experiments, particularly  those performed at low and ultralow temperatures, provide the most detailed  information about intermolecular interactions \cite{Kim:15,Wang:13b,Perreault:17,Vogels:15}. Molecular  collisions and chemical reactions at ultralow temperatures determine the stability of ultracold molecular gases, with ``good'' (elastic) collisions leading to desirable thermalization and ``bad'' (inelastic) collisions responsible for undesirable trap losses, which limits the stability of trapped molecules \cite{Krems:08,Carr:09,Balakrishnan:16,Bohn:17}. A detailed understanding of the quantum dynamics of binary collisions and chemical reactions  in ultracold environments may lead to novel ways to control them using, e.g., external electromagnetic fields \cite{Krems:08,Carr:09,Balakrishnan:16,Bohn:17} and/or quantum interference effects \cite{Devolder:21,Devolder:23}.

To model quantum scattering phenomena, one can use either time-independent or time-dependent (wavepacket) approaches \cite{Nyman:00,Althorpe:03,Zhang:16}.
Numerically exact quantum scattering simulations, in which the time-independent Schr\"odinger equation is solved directly via, e.g.,  coupled-channel methods or basis set expansion  techniques  \cite{Althorpe:03,Clary:08,Zhang:16},  are the "gold standard'' of chemical physics, similarly to the full configuration interaction or high-order coupled-cluster methods in quantum chemistry.
When such simulations are performed on classical computers, their computational complexity scales exponentially with the number of molecular degrees of freedom. As such, rigorous calculations are presently limited to systems containing a few atoms \cite{Zhang:16}.

To address the curse of dimensionality problem in quantum scattering simulations, QPE-based quantum algorithms have been developed  for chemical reactions  \cite{Kassal:08,Kassal:11}, atomic and molecular resonances \cite{Bian:21}, and nuclear scattering \cite{Roggero:2019,Du:2021}. While these algorithms achieve exponential speedup over their classical counterparts, they are designed for fault-tolerant quantum computers, rather than NISQ  devices \cite{McArdle:19}. Very recently, a hybrid VQE-based quantum algorithm has been proposed for solving the real-time chemical dynamics at low energies \cite{Lee:22}.
Yet, to our knowledge, no general-purpose quantum algorithm exists for solving the  time-independent quantum scattering problem for atoms and molecules on NISQ processors.

Here, we propose and implement
such an algorithm based on the $S$-matrix version of the Kohn variational principle (KVP) \cite{Zhang:88,Zhang:89}.
Our quantum KVP (Q-KVP) algorithm is conceptually simple and has the ability to  handle a wide range of  elastic, and inelastic, and reactive scattering problems on an equal footing using square-integrable ($L^2$) basis set expansions.  The fundamental scattering $S$-matrix, which encodes all scattering observables, is computed in the KVP using only matrix multiplications and inversions. The most computationally intensive step of the algorithm  involves the inversion of a real symmetric  Hamiltonian matrix  in the basis of real-valued $L^2$ basis functions. In the Q-KVP algorithm, the inversion problem is solved using the  variational quantum linear solver (VQLS), a recently developed NISQ variational algorithm \cite{VQLS}. We illustrate the Q-KVP algorithm by applying it to nontrivial single-channel and multichannel  quantum scattering problems. We finally demonstrate  how the Q-KVP algorithm can be  scaled up to larger molecular collision  systems, which are currently beyond the capabilities of classical computers.

The key quantity in quantum scattering theory is the multichannel $S$-matrix with elements $S_{n_fn_i}$, which encode the transition amplitudes between the initial and final channels $n_i$ and $n_f$.

 In the framework of the KVP, the $S$-matrix element is obtained by extremizing  the expression \cite{Zhang:1988vy} 
\begin{equation}\label{SmatKVP}
S_{n_f n_i}(c_{l m n}) =\text{ext} [ c_{1n_f n_i} + i \langle \tilde{\psi}_{n_f}| \hat{H}-E| \tilde{\psi}_{n_i}\rangle ]
\end{equation}
 with respect to the parameters $\{c_{l m n}\} $ of the trial wavefunctions for the initial and final scattering channels 
 \begin{align}\label{ansatz} \notag
  \tilde{\psi}_{n_i}(R,\mathbf{r})&=-u_{0n_i}(R) \phi_{n_i}(\mathbf{r})+u_{1n_i}(R) \phi_{n_i}(\mathbf{r}) \\ \notag
  &+\sum_{l=2}^{N_l}\sum_{n=1}^{N} c_{l n_in} u_{ln}(R)\phi_{n}(\mathbf{r}), \\ \notag
    \tilde{\psi}_{n_f}(R,\mathbf{r})&=-u_{0n_f}(R) \phi_{n_f}(\mathbf{r})+u_{1n_f}(R) \phi_{n_f}(\mathbf{r}) \\
    &+\sum_{l=2}^{N_l}\sum_{n=1}^{N} c_{l n_fn} u_{ln}(R)\phi_{n}(\mathbf{r}),
  \end{align}
where $\hat{H}$ is the Hamiltonian of the system (see below), $E$ is the total energy, $R$ is the scattering or reaction coordinate, $\mathbf{r}$ are the internal coordinates of the colliding molecules (such as their vibrational and rotational modes), and $\phi_n(\mathbf{r})$ is an internal wavefunction.  The $L^2$ basis functions $u_{ln}(R), l=2,3,\ldots, N_l$ describe the collision complex  at short $R$, and thus  $u_{ln}(R)\to 0$ as $R\to \infty$. The total number of scattering channels   $N$ includes both open (energetically allowed) and closed (energetically forbidden) channels.

At large $R$, the last terms on the right-hand side of Eq.~\eqref{ansatz} vanish, but the continuum basis functions $u_{0n}(R)$ and  $u_{1n}(R)=u^*_{0n}(R)$ do not. We parametrize these functions as \cite{Zhang:1988vy} (in atomic units, where $ \hslash=1$)
   \begin{align}\label{U0l}\notag
   u_{0n}(R)&=f(R)e^{-ik_nR}v_n^{-1/2},\\ 
   u_{ln}(R)&=F_lR^{(l-1)}e^{-\gamma R} \quad (l=2,3, ..., N_l),
   \end{align}
   where $k_n=\sqrt{2\mu (E-E_{n})}$ is the asymptotic wavenumber in channel $n$ with  asymptotic energy $E_n$, $v_n=k_n/\mu$ is the asymptotic velocity, and $\mu$ is the reduced mass for the collision.

Equation~\eqref{U0l} ensures that our  trial wavefunctions \eqref{ansatz} behave properly  in the limit $R \to \infty$,  as linear combinations of incoming and outgoing waves (for open channels).  
   The purpose of the cutoff function $f(R)=1-e^{-\gamma R}$  in the first line of Eq.~\eqref{U0l} is to regularize  the continuum basis functions $u_{0n}(R)$   and $u_{1n}(R)=u_{0n}^*(R)$   at the origin, i.e., to ensure that $u_{0n}(R=0)=0$. In Eq.~\eqref{U0l}, $\gamma>0$ is a parameter for the bound-state basis functions $u_{ln}(R)$ with $l\ge 2$ and $F_l$ is a normalization factor. We note that the functions $u_{ln}(R)$  are not orthogonal, i.e., $\langle u_{ln} | u_{l'n}\rangle \ne \delta_{ll'}$ ($l,l'\ge 2$).

  By extremizing the expression in Eq.~(\ref{SmatKVP}) with respect to the variational parameters of the trial wavefunction (\ref{ansatz}), $\frac{\partial}{\partial c_l}[c_{1n_2 n_1} + i \langle \tilde{\psi}_{n_2}|H-E| \tilde{\psi}_{n_1}\rangle]=0$, and applying the L{\"o}wdin-Feshbach projection technique to separate the matrix operations involving the real and complex matrix elements,  we obtain the $S$-matrix as  \cite{Zhang:89}  
  \begin{equation}\label{SmatMultichannel}
  \mathbf{S}=i (\mathbf{B} - \mathbf{C}^T \mathbf{B}^{*-1}\mathbf{C} ),
  \end{equation}
   where $\mathbf{S}$, $\mathbf{B}$, and $\mathbf{C}$ are square matrices in the channel index $n$
    \begin{align}\label{BandC}\notag
    \mathbf{B} &=\mathbf{M}_{0,0}-\mathbf{M}_0^T \mathbf{M}^{-1}\mathbf{M}_0 \\
    \mathbf{C} &=\mathbf{M}_{1,0}-\mathbf{M}_0^{*T} \mathbf{M}^{-1}\mathbf{M}_0.
  \end{align}   
Here,  $\mathbf{M}_{0,0}$ and $\mathbf{M}_{1,0}$  have the dimensions $N_o \times N_o$, where $N_o$ is the number of open channels (those, for which $k_n^2 > 0$). These are small square matrices with elements  $(\mathbf{M}_{0,0})_{nn'}=\langle u_{0n}\phi_n|\hat{H}-E|u_{0n'}\phi_{n'}\rangle$ and $(\mathbf{M}_{1,0})_{nn'}=\langle u_{0n}^*\phi_n|\hat{H}-E|u_{0n'}\phi_{n'}\rangle$ with $n,n'=1,2,\ldots, N_o$, where the basis functions are defined in Eq.~(\ref{ansatz}).  We follow the convention of Ref.~\cite{Zhang:1988vy} in  assuming that the  continuum  basis functions $u_{0n}(R)$ and $u_{1n}(R)$  are  not complex conjugated in the bra vectors.
The  $(N_l-1)  N \times N_o$ complex rectangular matrix $\mathbf{M}_0$ is composed of the elements $(\mathbf{M}_{0})_{ln,n'}=\langle u_{ln}\phi_n|\hat{H}-E|u_{0n'}\phi_{n'}\rangle$ with  $n$ ranging from 1 to $N$ and $n'$ from 1 to $N_o$. Finally, $\mathbf{M}$ is a real symmetric  matrix  with dimension  $(N_l -1)  N \times (N_l -1)  N$  and elements $(\mathbf{M})_{ln,l'n'}=\langle u_{ln}\phi_n | \hat{H}-E | u_{l'n'}\phi_{n'}\rangle$.

 Because Eq.~\eqref{SmatMultichannel} involves the inversion and multiplication of small $N_o\times N_o$ matrices $\mathbf{B}$ and $\mathbf{C}$, it may seem that the $S$-matrix can be computed  efficiently on a classical computer. 
 However,  computing $\mathbf{B}$ and $\mathbf{C}$ using Eq.~\eqref{BandC}  requires the inversion of a large real symmetric  matrix $\mathbf{M}$, which is the main computational bottleneck in computing the $S$-matrix in classical KVP \cite{Zhang:1988vy,Zhang:88}.
 Here, we overcome this bottleneck by using the VQLS, a recently proposed hybrid quantum-classical algorithm for solving linear systems of equations \cite{VQLS}.


The central strategy of the VQLS algorithm is to variationally prepare a quantum state $\ket{x}$ satisfying $\mathbf{A}\ket{x}\propto \ket{b}$ or, equivalently, $ \mathbf{A}\ket{x}/\sqrt{\bra{x}\mathbf{A}^{\dagger}\mathbf{A}\ket{x}} \approx \ket{b}$, where $\mathbf{A}$ is a real symmetric matrix and $\ket{b}$ is a normalized version of a vector $\Vec{b}$. The Q-KVP method includes both real and complex types of quantum linear systems in Eqs.~\eqref{SmatMultichannel} and \eqref{BandC}: $\mathbf{B}^*\ket{x}=\ket{b}$ and $\mathbf{M}\ket{x}=\ket{b}$. The complex $\mathbf{B}$-matrix is  small in most scattering problems, and we neglect it here.



{Inverting the matrix} $\mathbf{M}$ amounts to solving a quantum linear systems problem, $\mathbf{M} \Vec{x}_k = \Vec{b}_k, k=1,2, \dots, 2^n$,
 where $n$ is the number of qubits, $\mathbf{M}$ is a real symmetric $2^n \times 2^n$ matrix, and $\Vec{b}_k$ is a unit vector of $\mathbb{R}^{2^n}$ such that $\Vec{b}_1=(1,0,0,\ldots,0)$,  $\Vec{b}_2=(0,1,0,\ldots,0)$, etc.
 Then, the vector {$\Vec{x}_k = \mathbf{M}^{-1} \Vec{b}_k$ forms  the $k$-th column of  $\mathbf{M}^{-1}$.} Using  Dirac's notation, $\Vec{x}_k \to \ket{x_k}$ and $\Vec{b_k}\to\ket{b_k}$, we can recast the matrix $\mathbf{M}^{-1}$ as

\begin{align} \notag 
 &\begin{pmatrix} \mel{b_0}{\mathbf{M}^{-1}}{b_0} & \mel{b_0}{\mathbf{M}^{-1}}{b_1} & \cdots & \mel{b_0}{\mathbf{M}^{-1}}{b_{2^n}} \\
\mel{b_1}{\mathbf{M}^{-1}}{b_0} & \mel{b_1}{\mathbf{M}^{-1}}{b_1} & \cdots & \mel{b_1}{\mathbf{M}^{-1}}{b_{2^n}} \\
\vdots & \vdots & \ddots & \vdots \\
\mel{b_{2^n}}{\mathbf{M}^{-1}}{b_0} & \mel{b_{2^n}}{\mathbf{M}^{-1}}{b_1} & \cdots & \mel{b_{2^n}}{\mathbf{M}^{-1}}{b_{2^n}}
\end{pmatrix} \\ \\
 &= \begin{pmatrix} \braket{b_0}{x_0} & \braket{b_0}{x_1} & \cdots & \braket{b_0}{x_{2^n}} \\
\braket{b_1}{x_0} & \braket{b_1}{x_1} & \cdots & \braket{b_1}{x_{2^n}} \\
\vdots & \vdots & \ddots & \vdots \\
\braket{b_{2^n}}{x_0} & \braket{b_{2^n}}{x_1} & \cdots & \braket{b_{2^n}}{x_{2^n}}
\end{pmatrix}.\label{eq:M-1}
\end{align}

Thus, we need  to solve a system of $2^n$ linear equations to get the complete representation of $\mathbf{M}^{-1}$.

In the framework of the VQLS algorithm \cite{VQLS}, we initially prepare four fundamental inputs: matrices $\mathbf{P}$, a gate $\mathbf{U}$, an ansatz $V(\theta)$ and a cost function $C$. First, we decompose $\mathbf{M}$ into a linear combination of unitary matrices $\mathbf{M}_l$.
  In general, we can choose $\mathbf{M}_l$ in the form of products of Pauli matrices $\mathbf{P}_l \in \{ \mathbf{I},\mathbf{X},\mathbf{Y},\mathbf{Z} \}^{\otimes n}$ 
\begin{equation}\label{Mexpansion}
 \mathbf{M} = \sum_l c_l \mathbf{P}_l,
\end{equation}
where the expansion coefficients are given by $c_l=(1/2^n) \Tr\left(\mathbf{M}\mathbf{P}_l\right)$.

\begin{figure}[t]
\centering
\includegraphics[width=0.5\textwidth]{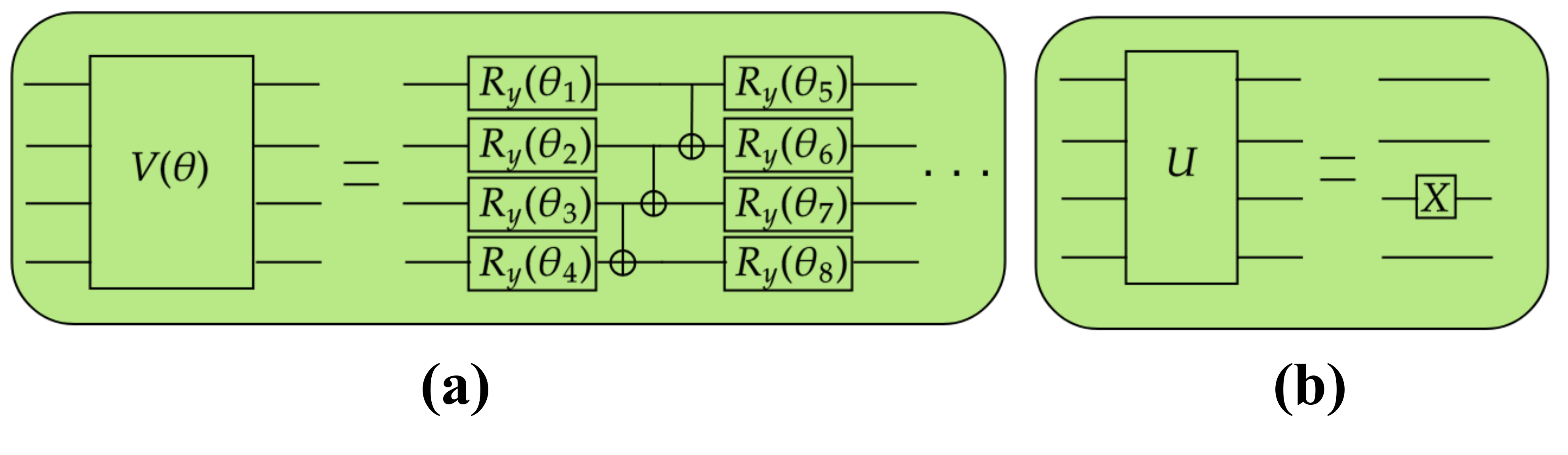}
\caption{Quantum circuits: (a) The ansatz $V(\theta)$ consists  of alternating layers of $Y$ gates and $CX$ gates. (b) The unitary $U$ is efficiently implemented with $X$ gates and $I$ gates.}
\label{fig:ansatz}
\end{figure}

 Second, we create a parametrized quantum circuit using a multiply layered hardware efficient ansatz $V(\theta)$ such that our trial solution is of the form $\ket{x(\theta)} = V(\theta) \ket{0}$, where  $\theta$ represents a set of variational parameters and {$\ket{0}=\ket{0}^{\otimes n}$ is the initial state of $n$ qubits} (see Fig.~\ref{fig:ansatz}). Each layer of the ansatz consists of single-qubit $Y$ gates and two-qubit  $CX$ gates. The number of layers is adjustable depending on the problem, which also represents the depth of the ansatz.


 Third, we prepare a quantum state $\ket{b}$ proportional to the vector $\Vec{b}$ using an efficient gate sequence $U$ shown in Fig.~\ref{fig:ansatz} (b), {which only includes the $X$ and $I$ gates.}
  Finally, to obtain an optimal $\theta$ to ensure $ \mathbf{M} \ket{x(\theta)}/\sqrt{\mel{x(\theta)}{\mathbf{M}^\dagger \mathbf{M}}{x(\theta)}} \approx \ket{b}$, we use the local version of the cost function in the VQLS algorithm \cite{VQLS}
\begin{equation}\label{cost_fun}
C =  1 - \frac{\left|\braket{b}{\Phi} \right|^2}{\mel{x(\theta)}{\mathbf{M}^\dagger \mathbf{M}}{x(\theta)} }.
\end{equation}
where $\ket{\Phi} = \mathbf{M} \ket{x(\theta)}$. Combining Eqs.~\eqref{Mexpansion} and \eqref{cost_fun}, the  cost function becomes
\begin{equation}
 C =  1 - \frac{\sum_{ll'} c_l c_{l'}^* \bra{0} U^\dagger \mathbf{P}_{l'} V(\theta) \ket{0} \bra{0} V(\theta)^\dagger \mathbf{P}_l^\dagger U \ket{0} }{\sum_{ll'} c_l c_{l'}^* \bra{0} V(\theta)^\dagger \mathbf{P}_{l'}^\dagger \mathbf{P}_l V(\theta) \ket{0}}
\end{equation}
Here, $U\ket{0}$ and $\ket{0}\bra{0}$ are replaced by  $ \ket{b}$ and $\mathbf{O} = \dfrac{1}{2}+\dfrac{1}{2n}\sum_{j=0}^{n-1}\mathbf{Z}_j$, respectively, in order to make the  estimation of expectation values easier, and the Pauli $\mathbf{Z}$ operator is locally implemented on the $j$-th qubit ($j=0,1,\ldots,n-1$).

\begin{figure}[t]
\centering
\includegraphics[width=0.9\linewidth]{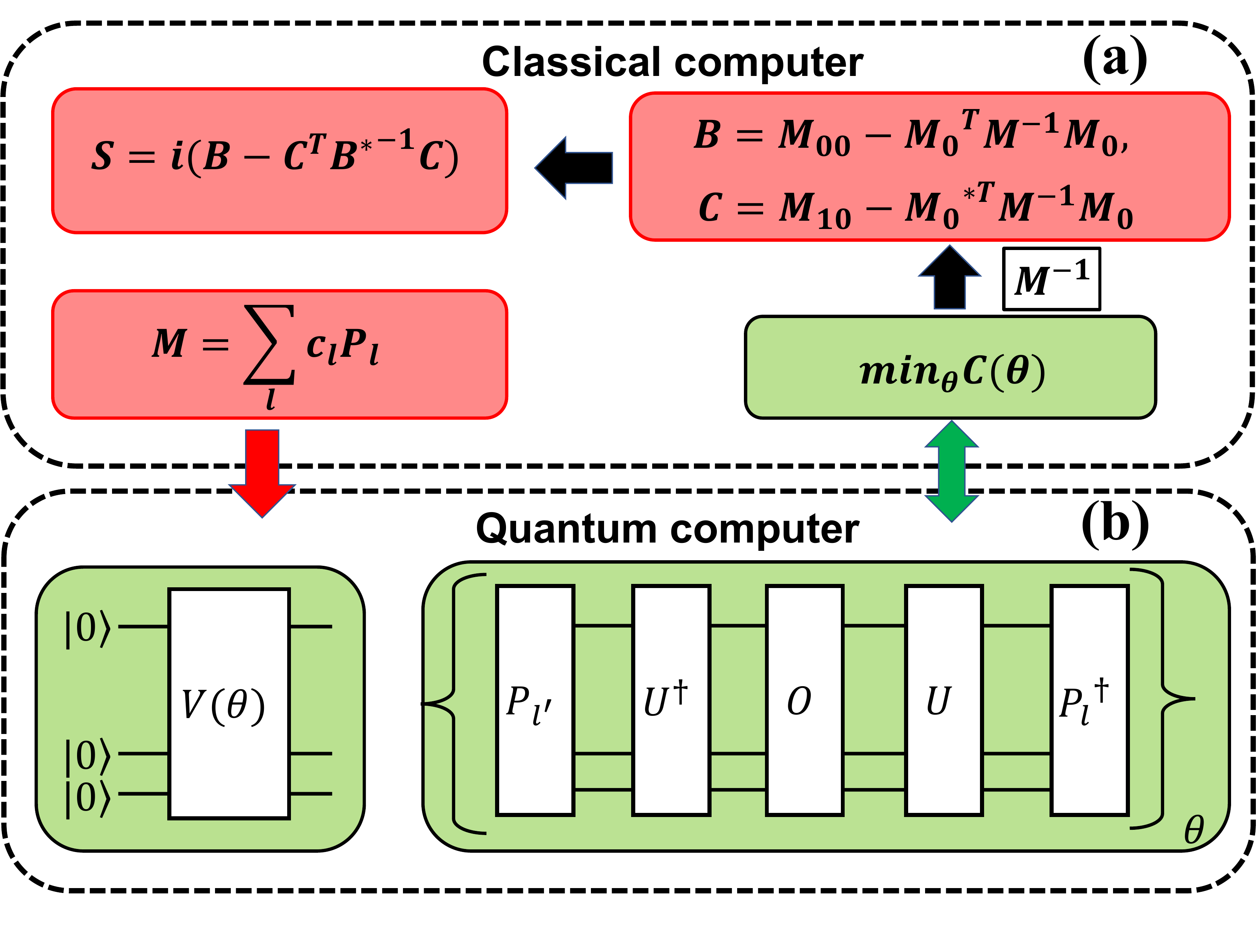}
\caption{Schematic diagram of the hybrid Q-KVP algorithm.}
\label{fig:diagram}
\end{figure}

By minimizing the cost function, we  obtain the optimal variational parameters $\theta$ and the optimal solution  $\ket{x}=\ket{x(\theta)}= V(\theta) \ket{0}$,  which satisfies  $ \mathbf{M} \ket{x(\theta)}/\sqrt{\mel{x(\theta)}{\mathbf{M}^\dagger \mathbf{M}}{x(\theta)}} \approx \ket{b}$. As the ansatz $V(\theta)$ is unitary, $\ket{x(\theta)}$ has unit norm.
 The vector $\Vec{x}=\mathcal{N}_q\ket{x(\theta)}$ has the quantum norm $\mathcal{N}_q=\norm{\ket{b}}/\norm{\mathbf{M} \ket{x(\theta)}}$, which is expected to be close to the exact classical norm $\mathcal{N}_c=\norm{\Vec{x}}$ (the difference between $\mathcal{N}_q$ and $\mathcal{N}_c$ is quantified by the fidelity, see below). Finally,  we assemble  the matrix $\mathbf{M}^{-1}$ {from the solution vectors}. The sign of $\vec{x}$ is determined by requiring $\mathbf{M }\mathbf{M}_q^{-1}=1$. The abovementioned steps  of the Q-KVP method are illustrated in Fig.~\ref{fig:diagram}. 


We begin by applying the  Q-KVP algorithm to a model single-channel quantum scattering problem  described by the Hamiltonian  
 \begin{equation}\label{POT-1D}
 H=-\dfrac{1}{2 \mu}\dfrac{d^2}{ dR^2} +V(R),
 \end{equation}
where  $V(R)=-e^{-R}$ \cite{Zhang:1988vy}. To this end, we define a single-channel ansatz 
 \begin{equation}\label{wf-1D}
\psi_t(R)=-u_0(R)+u_1(R)+\sum_{l=2}^{N_l}c_lu_l(R),
 \end{equation}
where $k=\sqrt{2\mu E}$ is the wavenumber, $E$ is the total energy, $v=k/\mu$ is the asymptotic velocity, $u_0=f(R)e^{ikR}v^{-1/2}$, $f(R)$ is the cutoff function defined above, and $u_1(R)=u_0(R)^*$. We take the bound-state basis functions to be $u_l(R)=F_lR^{l-1}e^{-\gamma R}$ ($l=2,3, ..., N_l$), and define the matrix elements 
 \begin{align}\label{Ul}\notag
  M_{00}&=\langle u_0|H-E|u_0\rangle,    M_{10}=\langle u_1|H-E|u_0\rangle;\\ \notag
   (\mathbf{M}_{0})_l&=\langle u_l|H-E|u_0\rangle,     (\mathbf{M})_{ll'}=\langle u_l|H-E|u_{l'}\rangle,   
   \end{align}
where $M_{00}$ and $M_{10}$ are complex scalars, $\mathbf{M}_0$ is a complex vector, and $\mathbf{M}$ is a real symmetric square matrix (see above). Substituting these expressions into Eq.~(\ref{BandC}),  we obtain the $S$ matrix element, which, for single-channel scattering, is a complex number of unit magnitude ($|S|^2=1$).
 The bound-state basis functions $u_{l}(R)$ are parametrized by $\gamma=1.5$ and we take $\mu=1$ a.u.

 \begin{figure}[t]
	\centering
	\includegraphics[width=8 cm]{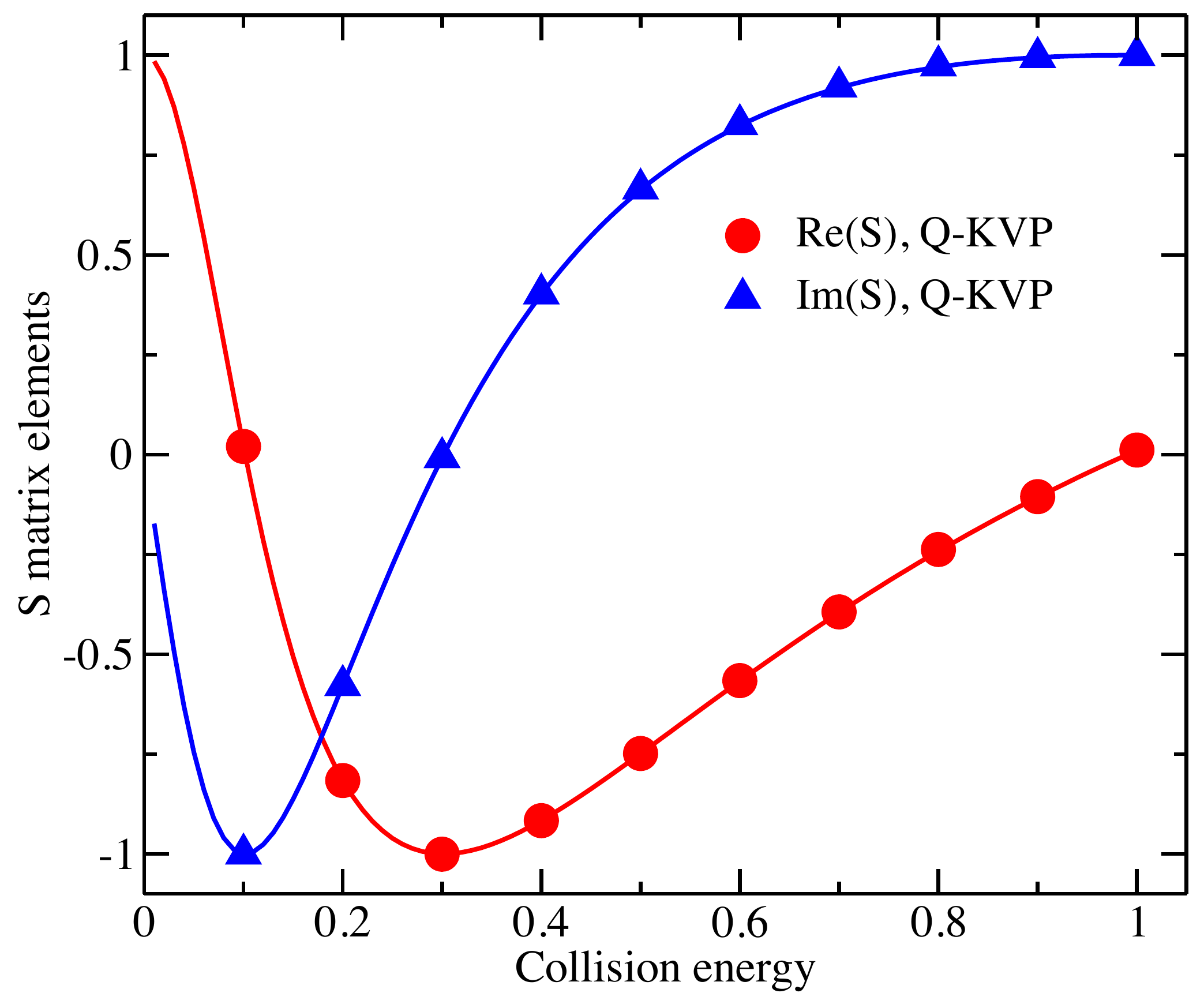}
	\caption{The real and imaginary parts of the $S$-matrix  plotted as a function of collision energy. Symbols -- Q-KVP results calculated with $N_l=2$. Solid lines -- benchmark  results computed using the CC method on a classical computer.
	The error bars on the quantum results are much smaller than the size of the symbols.  }
	\label{fig:1D-S}
\end{figure}

To compute the $S$-matrix using the Q-KVP algorithm, we start by inverting the $\mathbf{M}$-matrix in the framework of the VQLS algorithm on the Qiskit platform. The quality of the unit vectors $\ket{x_q}=\ket{x(\theta)}$ obtained using VQLS is measured by the fidelity
\begin{equation}\label{fidelity}
\mathcal{F}_{qc}=|\bra{x_q}\ket{x_c}|^2,
\end{equation}
which is a squared overlap between  $\ket{x_q}$ and the exact result, $\ket{x_c}=\Vec{x}/\mathcal{N}_c$, obtained by a classical inversion method. We obtain  fidelities very close to 1 using an ansatz with two layers as discussed in more detail in the Supporting Information (SI)\cite{SI}. 




Figure \ref{fig:1D-S} shows the collision energy dependence of the real and imaginary parts of the $S$-matrix calculated using our Q-KVP approach. The quantum results are in excellent agreement with the benchmark values obtained by direct numerical integration of the Schr{\"o}dinger equation on a classical computer using the numerically exact coupled channel (CC) method \cite{Johnson:73} over the whole range of collision energies.
As a point of reference, at $k= 0.55$,  our Q-KVP calculations give $S=-0.65714 + i0.75633$ for $N_l=2$, which compares favorably with the exact CC result ($S=-0.65769 + i0.75328$). Increasing the number of basis functions to 4 changes the imaginary part of the $S$-matrix element by only 0.4\% (and its real part by much less), 
 indicating good convergence of the Q-KVP results.



To explore the applicability of the Q-KVP algorithm to more realistic (and more complex) multi-channel quantum scattering problems, we apply it to solve a two-dimensional Secrest-Johnson (SJ) model   \cite{SJ-model,Stechel:1978}, which describes collinear collisions of a diatomic molecule with a structureless atom. The molecule is approximated by a harmonic oscillator, and a model two-dimensional (2D) potential  function is  used to describe the atom-molecule  interaction (see below). Because of these features,  the SJ model is significantly more complex  than the one-channel problem considered  in the previous section, and it continues to serve as a benchmark for testing new methods for solving CC equations of quantum scattering theory \cite{Manolopoulos:95}.

The Hamiltonian of the SJ model  is, in reduced coordinates \cite{SJ-model,Stechel:1978} 
 \begin{equation}\label{schordinger-2D}
H=-\dfrac{1}{2\mu}\dfrac{\partial^2}{\partial R^2} + V(R,r) -\dfrac{1}{2m}\dfrac{\partial^2}{\partial r^2}+\dfrac{1}{2}r^2,
 \end{equation}
 where $R$ is the scaled distance between the atom and the molecule's center of mass,  $r$ is the scaled internuclear distance in the molecule, $\mu$ is proportional to the reduced mass of the atom-molecule system,  and we use the values $m=1$ and $\mu=0.6667$, which correspond to He~+~H$_2$ collisions \cite{SJ-model,Stechel:1978}.
The Hamiltonian includes the kinetic energy of the atom relative to the molecule (the first term), the atom-molecule interaction potential (the second term), and the vibrational energy of the diatomic molecule (the third and the fourth terms).


The atom-molecule interaction potential $V(R,r)$ in the SJ model has the form
 \begin{equation}\label{POT-2D}
V(R,r)=Ae^{-\alpha R + \beta r},
 \end{equation}
where  $A$ is a constant, which defines the classical turning point for the collision, and the parameters  $\alpha$ and $\beta$ characterize the exponential decay of the potential with $R$ and its dependence on the diatomic stretching coordinate $r$.  To parameterize the interaction potential, we choose the values $A=10$, $\beta=2$, and $\alpha =0.3$.
The scattering wavefunction $\ket{\psi_t}$ satisfies the time-independent Schr{\"o}dinger equation $H\ket{\psi_t}=\dfrac{1}{2}E\ket{\psi_t}$, where  the total energy $E$ is expressed in units of the zero-point vibrational energy of the diatomic molecule \cite{SJ-model}. 


 We choose the Q-KVP ansatz for $\ket{\psi_t}$ in the form of Eq.~(\ref{ansatz}) with the transitional wavefunctions given by  Eq.~(\ref{U0l}) with   $\gamma =0.5$.
   The internal wavefunctions $\phi_n(r)$  are the vibrational wavefuctions  of the harmonic oscillator $\phi_v(r)$, where $v$ is the vibrational quantum number.
 Because $V(R,r)\to 0$ as $R\to\infty$,  the threshold energies of the vibrational channels are given by 
the energy levels of the one-dimensional harmonic oscillator, $\epsilon_v=v+1/2$.
 As noted above, the dimension of the  $S$ matrix is equal to the number of open channels, $N_o$.
 To avoid the overcompleteness problem caused by non-orthogonality of the basis functions $|u_{ln}\phi_{n}\rangle$, we transform the matrix $\mathbf{M}$ to an orthonormal basis as described in the SI \cite{SI}. The transformed matrices have dimensions $N_q  \times N_q $ for $\bf{M}$ and $ N_q  \times N_o$ for $\mathbf{M}_0$, where $N_q$ is the number of orthogonalized basis functions. 
 
 

 \begin{center}
\begin{table*}[t]
\renewcommand{\arraystretch}{1.2} \addtolength{\tabcolsep}{5 pt}
\begin{tabular}{ccccccc}
\hline \hline
$k$ &$\mathcal{F}^1$ &$\mathcal{F}^2$&$\mathcal{F}^3$& $\tau^1$ (s) &$\tau^2$ (s) &$\tau^3$ (s) \\
\hline 
1 & 0.1588 &0.5834&0.9999&63.8&1355.4&2699.6\\
2 & 0.6446 &0.9988&0.9999&84.5&1411.6&2657.3\\
3 & 0.8731 &0.9884&0.9999&55.3&1865.9&3267.6\\
4 & 0.0925 &0.9785&0.9999&80.7&1681.9&3872.4\\
5 & 0.0101&0.9999&0.9999&41.9&2155.7&2536.6\\
6 & 0.1232&0.9866&0.9999&105.4&2104.1&2227.5\\
7 & 0.9264&0.9967&0.9999&69.6&1787.7&3197.0\\
8 & 0.0058&0.9911&0.9999&55.7&1936.7&2838.1\\
\hline \hline
\end{tabular}
\caption{The  fidelity $\mathcal{F}$ and  computing  time $\tau$ of inverting  the $\bf M$ matrix for $E=3.8$ for the SJ problem with two open channels. The superscript $d$ of $\mathcal{F}^{(d)}$ and  $\tau^{(d)}$  indicates the number of layers in the ansatz. The dimension of the $\bf M$ matrix is $8\times 8$. }
\label{tab:2D-2ch}
\end{table*}
\end{center}

 Table~\ref{tab:2D-2ch} lists the fidelities for inverting the $\mathbf{M}$-matrix using VQLS in a basis of 8 orthogonalized functions. 
 The fidelity $\mathcal{F}^{(d)}$ is computed for ansatzes in Fig.~\ref{fig:ansatz}(a) with varying numbers of layers $d$ (for depth).
 We observe that  a single-layer ansatz  cannot provide a consistently good fidelity for all columns $k$ of $\mathbf{M}^{-1}$. The values of $\mathcal{F}^{(1)}$ vary from 0.0925 to 0.9264. Increasing the depth of the ansatz leads to a dramatic improvement in inversion fidelity;  values of $\mathcal{F}^{(3)}=0.9999$ are obtained using only three layers. A downside of using the $d\ge 3$ ansatses  is that their  computational time  increases rapidly. As shown in Table~\ref{tab:2D-2ch}, the computational cost of $\tau^{(3)}$ is nearly 50 times longer than that of $\tau^{(1)}$. Thus, inverting a general (i.e., non-sparse) $2^n \times 2^n$ $\mathbf{M}$  matrix using the VQLS algorithm on the Qiskit platform is very costly  for $n\geq 4$. 
This is because for a general, non-sparse matrix $\mathbf{M}$ the number of terms in its  expansion in products of Pauli matrices (\ref{Mexpansion}) grows exponentially with  matrix size. In order to avoid the explosive scaling, one should ensure that the number of Pauli terms grows only polynomially with system size. This important question is addressed below.

  \begin{figure}[t]
	\centering
	\includegraphics[width=8cm]{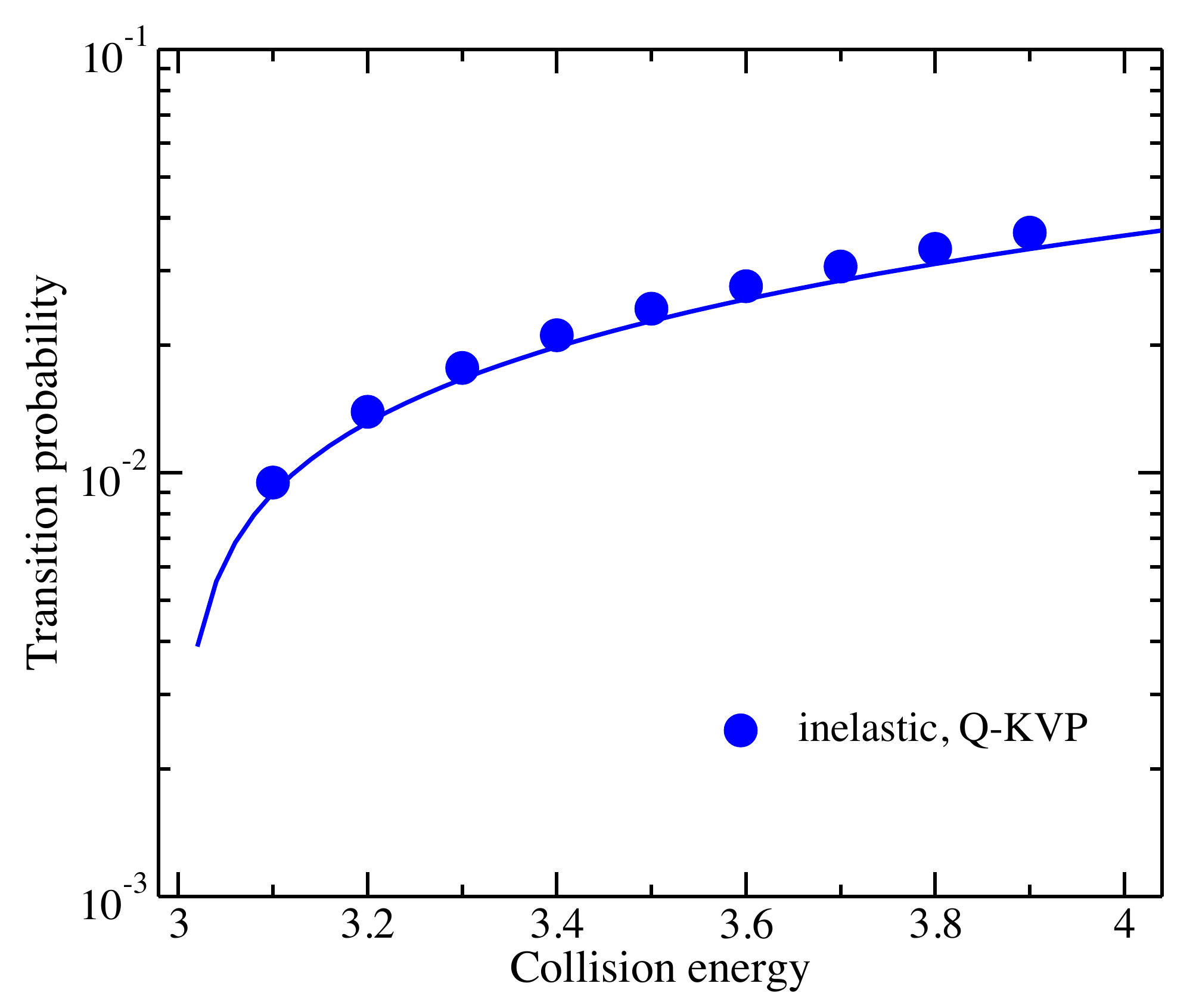}
	\caption{ The transition probabilities of the elastic and the inelastic collisions plotted as a function of collision energy for the case of two open channels ($v=0,1$). The size of the $\mathbf{M}$ matrix is $8 \times 8$. Solid symbols -- Q--KVP results.
	Solid lines -- reference results computed using exact CC.}
	\label{fig:2D-R}
\end{figure}
  
 Figure~\ref{fig:2D-R} compares the quantum and classical  transition probabilities for collision-induced vibrational relaxation ($|S_{10}|^2$). We consider the case of a diatomic molecule initially in the $v=1$ vibrational state colliding with a spherically symmetric atom, where the molecule can either scatter elastically or undergo vibrational relaxation to the $v=0$ ground state. The Q-KVP calculations use the $\mathbf{M}$ matrix in a basis of 8 orthogonalized basis functions, each of which is expanded in 12 primitive basis functions $|u_{ln}\phi_n\rangle$ (see the SI\cite{SI}).
The quantum probabilities are in very good agreement with exact CC calculations at all collision energies, validating our Q-KVP approach. In particular, the monotonic increase of the transition probability with collision energy is well reproduced by quantum calculations.

 \begin{figure}[t]
	\centering
	\includegraphics[width=8 cm]{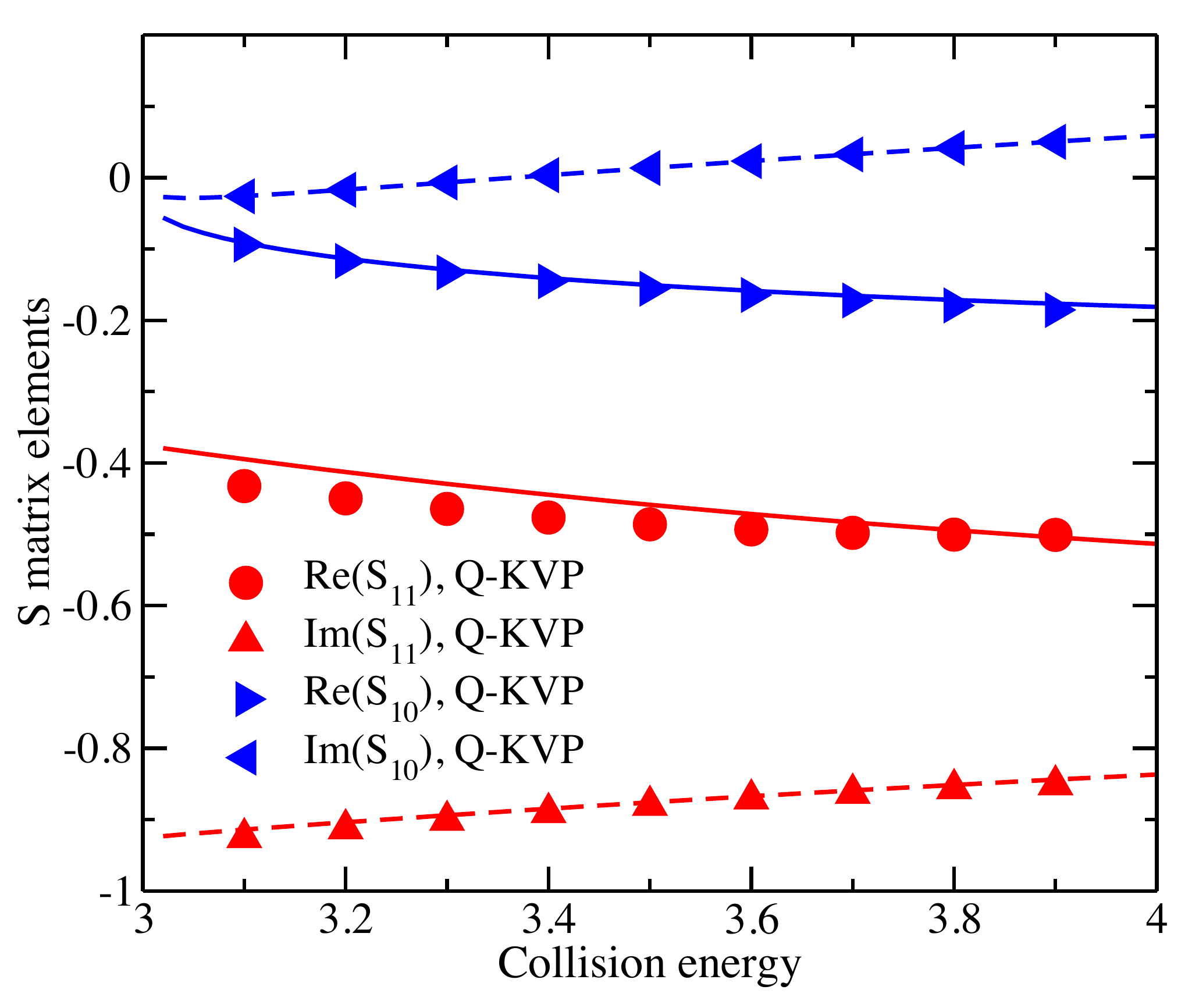}
	\caption{ The real and imaginary parts of the S-matrix plotted as a function of collision energy for the SJ model. Two  channels are open ($v=0,1$),  and the size of the $\mathbf{M}$ matrix is $8 \times 8$. Symbols -- Q--KVP results, solid lines -- reference results computed using exact CC. } 
	\label{fig:2D-S}
\end{figure}

 Having compared the inelastic transition probabilities, we now turn to the underlying complex transition amplitudes.
 Figure \ref{fig:2D-S} shows the real part and imaginary parts of the elastic and inelastic amplitudes ($S_{11}$ and $S_{10}$) obtained from our Q-KVP calculations. 
  As in the case of transition probabilities (see Fig.~\ref{fig:2D-R}), we observe excellent agreement between the Q-KVP and exact CC inelastic $S$-matrix elements at all collision energies.
However, the elastic S-matrix element ($Re(S_{11})$) deviates slightly from the exact CC result at lower energies. This is likely due to the limited size of our basis set, which contains 8 orthogonalized basis functions.

{\it To demonstrate the scalability of the Q-KVP algorithm}, ({\it i.e.} its ability to solve  large-scale  molecular scattering problems with exponential speedup over classical algorithms) we need to find a representation of the scattering Hamiltonian (or, equivalently, of $\mathbf{M}$)  that is $k$-local, i.e., with each $\mathbf{P}_l$ term in Eq. (\ref{Mexpansion})  acting in a non-trivial way on no more than $k$ qubits \cite{Cao:19}. If this condition is met, the number of Pauli terms  will grow polynomially with the size of the system, enabling efficient digital quantum simulation 
\cite{Cao:19,McArdle:19,McArdle:20}.
To this end, consider the Cartesian reaction path Hamiltonian that describes a wide range of molecular quantum  scattering processes, including inelastic collisions, quantum reactive scattering, and isomerization \cite{Ruf:88,Petkovic:03,Giese:05,Giese:05b}
   \vspace{-5pt}
\begin{equation}\label{Hrxp}
  \hat{H}= \sum_{i} \frac{P^2_i}{2M_i} + \sum_{i} \left[ \frac{p^2_i}{2m_i} + \frac{1}{2}m_i\omega_i (y_i-y_{i_0})^2 \right] + V(\mathbf{R},\mathbf{y}),
     \vspace{-7pt}
  \end{equation}
  where the Cartesian coordinates $\mathbf{R}=(R_1,R_2,\ldots)$ describe the large-amplitude motion of molecular fragments responsible for the chemical reaction or inelastic scattering, whereas the coordinates $\mathbf{y}$  describe the majority of molecular degrees of freedom that only exhibit small (nearly harmonic) displacements described by the  Hamiltonians $\hat{H}_i=\frac{p^2_i}{2m_i} + \frac{1}{2}m_i\omega_i (y_i-y_{i_0})^2 $. Accordingly, the adiabatic PES  can be expanded in these displacements \cite{Miller:80,Ruf:88} (assuming  the validity of the Born-Oppenheimer approximation)
    \vspace{-12pt}
  \begin{align}\label{Vrxp}\notag
 V(\mathbf{R},\mathbf{y}) &=  V(\mathbf{R},\mathbf{y}_0) + \sum_{i=1}^{N_M} K_{i}^{(1)}(\mathbf{R}) (y_i-y_{i_0}) \\
 +&  \sum_{i\ne j}^{N_M} K_{ij}^{(2)}(\mathbf{R}) (y_i-y_{i_0}) (y_j-y_{j_0}) + \ldots,
   \vspace{-7pt}
   \end{align}
where $V(\mathbf{R},\mathbf{y}_0)$ is the value of the interaction potential at the reference geometry with all   vibrational degrees of freedom frozen at their equilibrium positions $y_{i_0}$, $N_M$ is the number of vibrational modes, and ${K}^{(n)}_{i_1i_2\cdots i_n}$ are the expansion coefficients given by the $n$-th order derivatives of the interaction potential with respect to $y_{i_1}, y_{i_2}\ldots, y_{i_n}$, {\it i.e.} $K_{ij}^{(2)}=\frac{1}{2}\frac{\partial^2 V(\mathbf{R},\mathbf{y})}{\partial y_i\partial y_j}|_{\mathbf{y}=\mathbf{y}_0}$.

To recast the Hamiltonian (\ref{Hrxp}) into a $k$-local form, we choose an orthonormal  direct-product basis set 
$|u_{l} \rangle| \phi^{(1)}_{v_1} \phi^{(2)}_{v_2} \ldots  \phi^{(N_M)}_{v_M}  \rangle$ composed of $N_R$
scattering basis functions $\langle R |u_{l}\rangle = u_{l} (R)$ and $N_v^{N_M}$  vibrational basis functions $|\phi^{(1)}_{v_1}\rangle$, $| \phi^{(2)}_{v_2} \rangle, \cdots  |\phi^{(N_M)}_{v_M} \rangle$  (with $N_v$ functions per each  vibrational mode). For simplicity, we will consider the case of a single scattering variable $R$, which describes inelastic scattering or chemical reactions dominated by a single reaction coordinate.  The operator $\hat{H}-E=\hat{H}(E)$ then becomes
    \vspace{-5pt}

 \begin{align}\label{Hrep} \notag
\hat{H}(E)& =  \sum_{l,l'} \sum_{\substack{ v_1,v_2,\ldots \\ v_1',v_2',\ldots} } (| u_{l} \rangle \langle u_{l'} |) |\phi^{(1)}_{v_1} \phi^{(2)}_{v_2} \ldots \rangle
\\ & \langle \phi^{(1)}_{v_1'} \phi^{(2)}_{v_2'}  \ldots   | 
\langle u_{l}| \langle \phi^{(1)}_{v_1} \phi^{(2)}_{v_2} \ldots | \hat{H}(E)| \phi^{(1)}_{v_1'} \phi^{(2)}_{v_2'}  \ldots \rangle |u_{l'} \rangle 
 \end{align}
   \vspace{-5pt}
with the matrix elements of the Hamiltonian   (\ref{Hrxp}) given by
  \begin{align}\label{Hmatrix_el}\notag
&\langle u_{l}| \langle \phi^{(1)}_{v_1} \phi^{(2)}_{v_2} \ldots | \hat{H}(E)| \phi^{(1)}_{v_1'} \phi^{(2)}_{v_2'}  \ldots \rangle |u_{l'}\rangle \\ \notag
& = \delta_{v_1 v_1'}\delta_{v_2 v_2'}\ldots [  \langle u_l | \frac{P^2}{2M} - E \\ \notag
&+V(R,\mathbf{y}_0) | u_{l'}\rangle ] + \delta_{ll'} \langle \phi^{(1)}_{v_1} \phi^{(2)}_{v_2} \ldots | \sum_i\hat{H}_i | \phi^{(1)}_{v_1'} \phi^{(2)}_{v_2'}  \ldots \rangle \\ \notag
&+ \sum_{i=1}^{N_M} \langle u_l | K_{i}^{(1)}({R}) |  u_{l'}\rangle \delta^{(i)}_{v_1v_2\ldots; v_1'v_2'\ldots}  \langle \phi^{(i)}_{v_i} | (y_i-y_{i_0}) |  \phi^{(i)}_{v_i'}\rangle \\ \notag
& + \sum_{i,j=1}^{N_M} \langle u_l | K_{ij}^{(2)}(R) |  u_{l'}\rangle \delta^{(ij)}_{v_1v_2\ldots; v_1'v_2'\ldots}  \\ 
&\langle \phi^{(i)}_{v_i} | (y_i-y_{i_0}) |  \phi^{(i)}_{v_i'}\rangle  \langle \phi^{(j)}_{v_j} | (y_j-y_{j_0}) |  \phi^{(j)}_{v_j'}\rangle + \cdots.
  \end{align}
where we have included all the terms up to the second order [higher-order anharmonic terms have  a similar   structure involving sums of products of the matrix elements of $ K_{i_1i_2,\cdots,i_n}^{(n)}(R)$]. In Eq. (\ref{Hmatrix_el}), $\delta^{(i)}_{v_1,v_2\ldots; v_1'v_2'\ldots}$ denotes the   product of $N_M-1$  Kronecker delta symbols $\delta_{v_1v_1'}\delta_{v_2 v_2'}\ldots$, which does not include $\delta_{v_iv_i'}$ and $\delta^{(ij)}_{v_1,v_2\ldots; v_1'v_2'\ldots}$ stands for the product of $N_M-2$ Kronecker delta symbols, which does not include   $\delta_{v_iv_i'}$ and $\delta_{v_jv_j'}$.  
 Combining Eqs. (\ref{Hrep}) and (\ref{Hmatrix_el}) we obtain the Pauli decomposition of the scattering Hamiltonian
     \vspace{-9pt}
  \begin{multline}\label{terms1to4Paulis}
\hat{H}-E= 
\mathbf{H}^{(R)}\otimes \mathbf{1}^{(v_1)}  \otimes   \cdots
+  \mathbf{1}^{(R)}  \otimes \sum_{i=1}^{N_M} \biggl{[}  \cdots  \otimes \mathbf{1}^{v_{(i-1)}}  \otimes \mathbf{E}^{(v_i)}\otimes \mathbf{1}^{(v_{i+1})} \otimes   \cdots\biggr{]} \\
+  \sum_{i=1}^{N_M} \biggl{(} \sum_{ \substack{ i_1,i_2,\ldots \\  \alpha_1,\alpha_2,\ldots } }k^{(1) i_1i_2\cdots}_{\, i,\,\alpha_1\alpha_2\cdots} \sigma^{i_{1}}_{\alpha_1}  \sigma^{i_{2}}_{\alpha_2} \cdots \biggr{)}
  \otimes \mathbf{1}^{(v_1)} \otimes  \cdots \otimes \mathbf{y}_i^{(v_i)} \otimes \cdots \\
+ \sum_{ \substack{ i,j=1 \\   i\ne j } }^{N_M} \biggl{(} \sum_{ \substack{ i_1,i_2,\ldots\\  \alpha_1,\alpha_2,\ldots  } } k^{(2) i_1i_2\cdots}_{\, ij,\,\alpha_1\alpha_2\cdots} \sigma^{i_{1}}_{\alpha_1}  \sigma^{i_{2}}_{\alpha_2} \cdots \biggr{)}
  \otimes \mathbf{1}^{(v_1)}  \otimes  \cdots \otimes \mathbf{y}_i^{(v_i)} \otimes  \cdots \otimes \mathbf{y}_j^{(v_j)} \otimes  \cdots
  \\
  + \sum_{i=1}^{N_M} \biggl{(} \sum_{ \substack{ i_1,i_2,\ldots\\  \alpha_1,\alpha_2,\ldots  } }k^{(2) i_1i_2\cdots}_{\, ii,\,\alpha_1\alpha_2\cdots} \sigma^{i_{1}}_{\alpha_1}  \sigma^{i_{2}}_{\alpha_2} \cdots \biggr{)}
  \otimes \mathbf{1}^{(v_1)} \otimes  \cdots \otimes (\mathbf{y}_i^{2})^{(v_i)} \otimes \cdots \\
  + \text{Anharmonic terms},
 \end{multline}
where  $\mathbf{1}_R$ and $\mathbf{1}^{(v_i)}$ are unit operators in the scattering  space spanned by $N^{(R)}$ basis functions $|u_l\rangle$, and in  the Hilbert space of the $v_i$-th vibrational mode spanned by $N_v$ vibrational basis functions  $|\phi_{v_i}^{(i)}\rangle$,
 $\mathbf{H}^{(R)}=  \sum_{ \substack{ i_1,i_2,\ldots\\  \alpha_1,\alpha_2,\ldots  } }u^{i_1i_2\cdots}_{\alpha_1\alpha_2\cdots} \sigma^{i_{1}}_{\alpha_1}  \sigma^{i_{2}}_{\alpha_2} \cdots $ is the matrix representation of the first  term of the Hamiltonian (\ref{Hmatrix_el}) expanded in Pauli matrices acting on $n_S=\log(N_R)$ ``scattering'' qubits $\alpha_1$, $\alpha_2,\ldots$.  Further, $\mathbf{E}^{(v_i)}=\text{diag}(E^{(i)}_{0},E_1^{(i)},\ldots)$ are the diagonal matrices of vibrational energies of the $i$-th mode, and   $k^{(i) i_1i_2\cdots}_{i,\,\,\alpha_1\alpha_2\cdots}$ and  $ k^{(2) i_1i_2\cdots}_{\, ij,\,\alpha_1\alpha_2\cdots}$ are the expansion coefficients   in the Pauli decomposition (\ref{Mexpansion}) of the matrices $(\mathbf{K}^{(i)})_{ll'} =\langle u_l | K_{i}^{(1)}(R) |  u_{l'}\rangle$ and  $(\mathbf{K}^{(ij)})_{ll'} =\langle u_l | K_{ij}^{(2)}(R) |  u_{l'}\rangle$. The number of terms in these expansions generally scales as $O(N_T^{R})=O(2^{N_R})$. This does not pose a fundamental problem, however, because the number of reaction coordinates typically ranges from one to three for most chemical reactions of interest  \cite{Ruf:88} regardless of system size. {Importantly, therefore, the presence of a large number of terms in the Pauli expansions  of $\mathbf{H}^{(R)}$, $\mathbf{K}^{(i)}$, and $\mathbf{K}^{(ij)}$  does not affect the overall scaling with system size of the number of terms in Eq. (\ref{terms1to4Paulis}), which is determined instead by the much more numerous vibrational modes (see below).}
 Finally, the  $N_v\times N_v$ matrices  $ (\mathbf{y}^n_i)^{(v_i)}$ in Eq. (\ref{terms1to4Paulis}) are given by 
 $\{(\mathbf{y}_i^n)^{(v_i)}\}_{v_iv_i'}=\langle \phi^{(i)}_{v_i} | (y_i-y_{i_0})^n |  \phi^{(i)}_{v_i'}\rangle$.  
  
 To encode vibrational modes into the ``vibrational qubits'' labeled $(v_1), (v_2), \ldots$, we will use a compact mapping, which requires $n_v=\log N_v$ qubits per mode \cite{McArdle:19}, where the Pauli  decompositions of  matrices ${\mathbf{E}}^{(i)}$ and $(\mathbf{y}^n_i)^{(v_i)}$ contain only $O(N_v^2)$ and $O(N_v^{2n})$ terms, respectively.  
{This implies that the total number of terms in the Pauli expansion of the scattering Hamiltonian (\ref{terms1to4Paulis}) scales polynomially as $O(N_T^{R})\times N_M^2 \times O(N_v^{4})$ with the number of vibrational modes $N_M$ and the number of vibrational basis functions per mode $N_v$. The $k$-locality of the Cartesian reaction path Hamiltonian \eqref{Hrxp} suggests that is could be efficiently simulated on a quantum computer.}  If $m$-th order anharmonic terms are included in the expansion of the interaction potential (\ref{Vrxp}), the scaling becomes steeper [$O(N_T^{R})\times N_M^m \times O(N_v^{2m})$] but remains polynomial for any finite $m$.

  \begin{table}[b]
 \vspace{-8pt}
\caption{\label{tab:scaling}  Estimated number of qubits $n_q=n_S+N_M n_v$ and Pauli terms $N_P = N_T^{R} N_M^2 N_v^{4}$  for simulating collisions of ethylene (C$_2$H$_4$), benzene (C$_6$H$_6$) and naphthalene (C$_{10}$H$_8$) molecules. The basis set includes 4 scattering states ($N_R=2,N_T^R =16$)  encoded in 2 scattering qubits ($n_S=2$), and 2 vibrational basis functions ($N_v=2$) encoded into 1 qubit per vibrational mode ($n_v=1$). Eq.~(\ref{terms1to4Paulis}) is truncated at the harmonic level ($m =2$). Rotational degrees of  freedom  are neglected for simplicity since they are greatly outnumbered by vibrations already for medium-size molecules  \cite{Ruf:88}, and thus do not affect the scaling of scattering computations with system size.  }
\begin{tabular}{ccc}
  \hline
    \hline
{ Collision}   &  { $n_q$}   & { $N_P\times 10^3$}  \\
  \hline
C$_2$H$_4$ + C$_2$H$_4$  & 26 &   147.5  \\ 
C$_6$H$_6$ + C$_6$H$_6$  & 62 &   921.6  \\ 
C$_{10}$H$_8$ + C$_{10}$H$_8$   & 98 & 2,560    \\
\hline
\hline
\end{tabular}
\vspace{-10pt}
\end{table}
 
Table \ref{tab:scaling} lists the estimated number of qubits and Pauli terms required for the digital quantum simulation of quantum collision dynamics of medium-to-large molecules ethylene (C$_2$H$_4$), benzene (C$_6$H$_6$), and naphthalene (C$_{10}$H$_8$). Collisions involving such hydrocarbon molecules  play an important role in combustion  and possibly in the interstellar medium \cite{Kaiser:02}, and pose an intriguing ``sticking'' problem \cite{Patterson:10,Li:12,Piskorski:14} concerning the role of molecular vibrations in the formation of long-lived collision complexes.
The problem is described by a single scattering coordinate $R$, the  distance between the molecules' centers of mass. The number of small-amplitude vibrational modes of the collision complex is thus $N_M= (3N_a - 3) - 1 - N_{A}$,  where $N_a$ is the total number of nuclei,  and $N_A=8$ is the number of angular variables, including six Euler angles that specify the position of each nonlinear molecule with respect to the intermolecular axis $\mathbf{R}$ and the two angles that determine the orientation of $\mathbf{R}$ in the laboratory frame.  


 {\it In conclusion}, we proposed a hybrid classical-quantum algorithm for the solution of multichannel quantum scattering problems, which are ubiquitous not only in physical chemistry but also in atomic and molecular physics. The Q-KVP algorithm combines the $S$-matrix version of the Kohn variational principle (KVP) \cite{Zhang:1988vy} and the recently developed VQLS algorithm \cite{VQLS}  
to address the major computational bottleneck of the classical KVP algorithm, the inversion of a large real symmetric matrix  $\bf M$.
In our initial  Qiskit implementation of the Q-KVP methodology, we construct a $2^n \times 2^n$ matrix $\bf M$ using an orthogonalized finite basis set and invert it using hardware-efficient multilayered ansatzes. This  procedure gives accurate results for both single-channel and multichannel quantum scattering. We applied Q-KVP methodology to study vibrational energy transfer in collinear atom-diatom collisions using the archetypal SJ model. The vibrational transition probabilities computed using Q-KVP are in excellent agreement with exact CC calculations, demonstrating the validity and accuracy of  the Q-KVP methodology.
A current limitation of our approach is the large computational cost of  inverting $\mathbf{M}$-matrices of dimensions greater than 16 on the Qiskit platform, due to the the exponentially increasing number of terms in the Pauli matrix expansion of $\bf M$.   To address this limitation,  we propose a scenario of how the Q-KVP algorithm could be scaled up to larger inelastic collision or reactive scattering problems involving larger polyatomic molecules. This is achieved by explicitly expanding the few anharmonic (scattering) degrees of freedom in Pauli matrices  and using the compact mapping for the remaining harmonic degrees of freedom as done in previous work on the digital quantum simulation of vibrational energy levels \cite{McArdle:19}.

\section*{Supporting Information}
See the supporting information associated with this article for details of convergence tests, additional information on the implementation of the  VQLS, and for the basis set orthogonalization procedure.

\begin{acknowledgement}

We thank Robert Parrish for illuminating discussions.
 This work was supported by the NSF through the CAREER program (PHY-2045681).  A.F.I. is grateful to the Mitacs Globalink program and the Natural Science and Engineering Council (NSERC) of Canada for financial support.
 
\end{acknowledgement}

\bibliography{Master.bib}

\providecommand{\latin}[1]{#1}
\makeatletter
\providecommand{\doi}
  {\begingroup\let\do\@makeother\dospecials
  \catcode`\{=1 \catcode`\}=2 \doi@aux}
\providecommand{\doi@aux}[1]{\endgroup\texttt{#1}}
\makeatother
\providecommand*\mcitethebibliography{\thebibliography}
\csname @ifundefined\endcsname{endmcitethebibliography}
  {\let\endmcitethebibliography\endthebibliography}{}
\begin{mcitethebibliography}{67}
\providecommand*\natexlab[1]{#1}
\providecommand*\mciteSetBstSublistMode[1]{}
\providecommand*\mciteSetBstMaxWidthForm[2]{}
\providecommand*\mciteBstWouldAddEndPuncttrue
  {\def\EndOfBibitem{\unskip.}}
\providecommand*\mciteBstWouldAddEndPunctfalse
  {\let\EndOfBibitem\relax}
\providecommand*\mciteSetBstMidEndSepPunct[3]{}
\providecommand*\mciteSetBstSublistLabelBeginEnd[3]{}
\providecommand*\EndOfBibitem{}
\mciteSetBstSublistMode{f}
\mciteSetBstMaxWidthForm{subitem}{(\alph{mcitesubitemcount})}
\mciteSetBstSublistLabelBeginEnd
  {\mcitemaxwidthsubitemform\space}
  {\relax}
  {\relax}

\bibitem[Cao \latin{et~al.}(2019)Cao, Romero, Olson, Degroote, Johnson,
  Kieferov{\'a}, Kivlichan, Menke, Peropadre, Sawaya, Sim, Veis, and
  Aspuru-Guzik]{Cao:19}
Cao,~Y.; Romero,~J.; Olson,~J.~P.; Degroote,~M.; Johnson,~P.~D.;
  Kieferov{\'a},~M.; Kivlichan,~I.~D.; Menke,~T.; Peropadre,~B.; Sawaya,~N.
  P.~D. \latin{et~al.}  Quantum Chemistry in the Age of Quantum Computing.
  \emph{Chem. Rev.} \textbf{2019}, \emph{119}, 10856--10915\relax
\mciteBstWouldAddEndPuncttrue
\mciteSetBstMidEndSepPunct{\mcitedefaultmidpunct}
{\mcitedefaultendpunct}{\mcitedefaultseppunct}\relax
\EndOfBibitem
\bibitem[McArdle \latin{et~al.}(2020)McArdle, Endo, Aspuru-Guzik, Benjamin, and
  Yuan]{McArdle:20}
McArdle,~S.; Endo,~S.; Aspuru-Guzik,~A.; Benjamin,~S.~C.; Yuan,~X. Quantum
  computational chemistry. \emph{Rev. Mod. Phys.} \textbf{2020}, \emph{92},
  015003\relax
\mciteBstWouldAddEndPuncttrue
\mciteSetBstMidEndSepPunct{\mcitedefaultmidpunct}
{\mcitedefaultendpunct}{\mcitedefaultseppunct}\relax
\EndOfBibitem
\bibitem[Arg{\"u}ello-Luengo \latin{et~al.}(2019)Arg{\"u}ello-Luengo,
  Gonz{\'a}lez-Tudela, Shi, Zoller, and Cirac]{ArguelloLuengo:19}
Arg{\"u}ello-Luengo,~J.; Gonz{\'a}lez-Tudela,~A.; Shi,~T.; Zoller,~P.;
  Cirac,~J.~I. Analogue quantum chemistry simulation. \emph{Nature}
  \textbf{2019}, \emph{574}, 215--218\relax
\mciteBstWouldAddEndPuncttrue
\mciteSetBstMidEndSepPunct{\mcitedefaultmidpunct}
{\mcitedefaultendpunct}{\mcitedefaultseppunct}\relax
\EndOfBibitem
\bibitem[Otten \latin{et~al.}(2022)Otten, Hermes, Pandharkar, Alexeev, Gray,
  and Gagliardi]{Otten:2022}
Otten,~M.; Hermes,~M.~R.; Pandharkar,~R.; Alexeev,~Y.; Gray,~S.~K.;
  Gagliardi,~L. Localized Quantum Chemistry on Quantum Computers. \emph{J.
  Chem. Theory Comput.} \textbf{2022}, \emph{18}, 7205--7217\relax
\mciteBstWouldAddEndPuncttrue
\mciteSetBstMidEndSepPunct{\mcitedefaultmidpunct}
{\mcitedefaultendpunct}{\mcitedefaultseppunct}\relax
\EndOfBibitem
\bibitem[Fan \latin{et~al.}(2021)Fan, Liu, Li, and Yang]{fan:2021}
Fan,~Y.; Liu,~J.; Li,~Z.; Yang,~J. Equation-of-Motion Theory to Calculate
  Accurate Band Structures with a Quantum Computer. \emph{J. Phys. Chem. Lett.}
  \textbf{2021}, \emph{12}, 8833--8840\relax
\mciteBstWouldAddEndPuncttrue
\mciteSetBstMidEndSepPunct{\mcitedefaultmidpunct}
{\mcitedefaultendpunct}{\mcitedefaultseppunct}\relax
\EndOfBibitem
\bibitem[Colless \latin{et~al.}(2018)Colless, Ramasesh, Dahlen, Blok,
  Kimchi-Schwartz, McClean, Carter, de~Jong, and Siddiqi]{Colless:2018}
Colless,~J.~I.; Ramasesh,~V.~V.; Dahlen,~D.; Blok,~M.~S.;
  Kimchi-Schwartz,~M.~E.; McClean,~J.~R.; Carter,~J.; de~Jong,~W.~A.;
  Siddiqi,~I. Computation of Molecular Spectra on a Quantum Processor with an
  Error-Resilient Algorithm. \emph{Phys. Rev. X} \textbf{2018}, \emph{8},
  011021\relax
\mciteBstWouldAddEndPuncttrue
\mciteSetBstMidEndSepPunct{\mcitedefaultmidpunct}
{\mcitedefaultendpunct}{\mcitedefaultseppunct}\relax
\EndOfBibitem
\bibitem[Parrish \latin{et~al.}(2019)Parrish, Hohenstein, McMahon, and
  Mart\'{\i}nez]{Parrish:19}
Parrish,~R.~M.; Hohenstein,~E.~G.; McMahon,~P.~L.; Mart\'{\i}nez,~T.~J. Quantum
  Computation of Electronic Transitions Using a Variational Quantum
  Eigensolver. \emph{Phys. Rev. Lett.} \textbf{2019}, \emph{122}, 230401\relax
\mciteBstWouldAddEndPuncttrue
\mciteSetBstMidEndSepPunct{\mcitedefaultmidpunct}
{\mcitedefaultendpunct}{\mcitedefaultseppunct}\relax
\EndOfBibitem
\bibitem[Leontica \latin{et~al.}(2021)Leontica, Tennie, and
  Farrow]{Leontica:21}
Leontica,~S.; Tennie,~F.; Farrow,~T. Simulating molecules on a cloud-based
  5-qubit IBM-Q universal quantum computer. \emph{Commun. Phys.} \textbf{2021},
  \emph{4}, 112\relax
\mciteBstWouldAddEndPuncttrue
\mciteSetBstMidEndSepPunct{\mcitedefaultmidpunct}
{\mcitedefaultendpunct}{\mcitedefaultseppunct}\relax
\EndOfBibitem
\bibitem[A.Yu.Kitaev()]{QPE}
A.Yu.Kitaev, Quantum measurements and the Abelian Stabilizer Problem.
  \emph{arXiv:quant-ph/9511026} \relax
\mciteBstWouldAddEndPunctfalse
\mciteSetBstMidEndSepPunct{\mcitedefaultmidpunct}
{}{\mcitedefaultseppunct}\relax
\EndOfBibitem
\bibitem[Nielsen and Chuang(2010)Nielsen, and Chuang]{Nielsen:10}
Nielsen,~M.~A.; Chuang,~I.~L. \emph{{\it Quantum Computation and Quantum
  Information: 10th Anniversary Edition}}; Cambridge University Press,
  2010\relax
\mciteBstWouldAddEndPuncttrue
\mciteSetBstMidEndSepPunct{\mcitedefaultmidpunct}
{\mcitedefaultendpunct}{\mcitedefaultseppunct}\relax
\EndOfBibitem
\bibitem[Pezz\`e and Smerzi(2021)Pezz\`e, and Smerzi]{pezze:21}
Pezz\`e,~L.; Smerzi,~A. Quantum Phase Estimation Algorithm with Gaussian Spin
  States. \emph{PRX Quantum} \textbf{2021}, \emph{2}, 040301\relax
\mciteBstWouldAddEndPuncttrue
\mciteSetBstMidEndSepPunct{\mcitedefaultmidpunct}
{\mcitedefaultendpunct}{\mcitedefaultseppunct}\relax
\EndOfBibitem
\bibitem[Peruzzo \latin{et~al.}(2014)Peruzzo, McClean, Shadbolt, Yung, Zhou,
  Love, Aspuru-Guzik, and O'Brien]{Peruzzo:14}
Peruzzo,~A.; McClean,~J.; Shadbolt,~P.; Yung,~M.-H.; Zhou,~X.-Q.; Love,~P.~J.;
  Aspuru-Guzik,~A.; O'Brien,~J.~L. A variational eigenvalue solver on a
  photonic quantum processor. \emph{Nat. Commun.} \textbf{2014}, \emph{5},
  4213\relax
\mciteBstWouldAddEndPuncttrue
\mciteSetBstMidEndSepPunct{\mcitedefaultmidpunct}
{\mcitedefaultendpunct}{\mcitedefaultseppunct}\relax
\EndOfBibitem
\bibitem[McClean \latin{et~al.}(2016)McClean, Romero, Babbush, and
  Aspuru-Guzik]{McClean:16}
McClean,~J.~R.; Romero,~J.; Babbush,~R.; Aspuru-Guzik,~A. The theory of
  variational hybrid quantum-classical algorithm. \emph{New J. Phys.}
  \textbf{2016}, \emph{18}, 023023\relax
\mciteBstWouldAddEndPuncttrue
\mciteSetBstMidEndSepPunct{\mcitedefaultmidpunct}
{\mcitedefaultendpunct}{\mcitedefaultseppunct}\relax
\EndOfBibitem
\bibitem[Romero \latin{et~al.}(2018)Romero, Babbush, McClean, Hempel, Love, and
  Aspuru-Guzik]{Romero:18}
Romero,~J.; Babbush,~R.; McClean,~J.~R.; Hempel,~C.; Love,~P.~J.;
  Aspuru-Guzik,~A. Strategies for quantum computing molecular energies using
  the unitary coupled cluster ansatz. \emph{Quantum Sci. Technol.}
  \textbf{2018}, \emph{4}, 014008\relax
\mciteBstWouldAddEndPuncttrue
\mciteSetBstMidEndSepPunct{\mcitedefaultmidpunct}
{\mcitedefaultendpunct}{\mcitedefaultseppunct}\relax
\EndOfBibitem
\bibitem[Preskill(2018)]{Preskill:18}
Preskill,~J. Quantum {C}omputing in the {NISQ} era and beyond. \emph{{Quantum}}
  \textbf{2018}, \emph{2}, 79--99\relax
\mciteBstWouldAddEndPuncttrue
\mciteSetBstMidEndSepPunct{\mcitedefaultmidpunct}
{\mcitedefaultendpunct}{\mcitedefaultseppunct}\relax
\EndOfBibitem
\bibitem[McArdle \latin{et~al.}(2019)McArdle, Mayorov, Shan, Benjamin, and
  Yuan]{McArdle:19}
McArdle,~S.; Mayorov,~A.; Shan,~X.; Benjamin,~S.; Yuan,~X. Digital quantum
  simulation of molecular vibrations. \emph{Chem. Sci.} \textbf{2019},
  \emph{10}, 5725--5735\relax
\mciteBstWouldAddEndPuncttrue
\mciteSetBstMidEndSepPunct{\mcitedefaultmidpunct}
{\mcitedefaultendpunct}{\mcitedefaultseppunct}\relax
\EndOfBibitem
\bibitem[Sparrow \latin{et~al.}(2018)Sparrow, Mart{\'\i}n-L{\'o}pez,
  Maraviglia, Neville, Harrold, Carolan, Joglekar, Hashimoto, Matsuda, O'Brien,
  \latin{et~al.} others]{sparrow:18}
Sparrow,~C.; Mart{\'\i}n-L{\'o}pez,~E.; Maraviglia,~N.; Neville,~A.;
  Harrold,~C.; Carolan,~J.; Joglekar,~Y.~N.; Hashimoto,~T.; Matsuda,~N.;
  O'Brien,~J.~L. \latin{et~al.}  Simulating the vibrational quantum dynamics of
  molecules using photonics. \emph{Nature} \textbf{2018}, \emph{557},
  660--667\relax
\mciteBstWouldAddEndPuncttrue
\mciteSetBstMidEndSepPunct{\mcitedefaultmidpunct}
{\mcitedefaultendpunct}{\mcitedefaultseppunct}\relax
\EndOfBibitem
\bibitem[Sawaya and Huh(2019)Sawaya, and Huh]{Sawaya:19}
Sawaya,~N. P.~D.; Huh,~J. Quantum Algorithm for Calculating Molecular Vibronic
  Spectra. \emph{J. Phys. Chem. Lett.} \textbf{2019}, \emph{10},
  3586--3591\relax
\mciteBstWouldAddEndPuncttrue
\mciteSetBstMidEndSepPunct{\mcitedefaultmidpunct}
{\mcitedefaultendpunct}{\mcitedefaultseppunct}\relax
\EndOfBibitem
\bibitem[Johnson \latin{et~al.}(2011)Johnson, Amin, Gildert, Lanting, Hamze,
  Dickson, Harris, Berkley, Johansson, Bunyk, Chapple, Enderud, Hilton, Karimi,
  Ladizinsky, Ladizinsky, Oh, Perminov, Rich, Thom, Tolkacheva, Truncik,
  Uchaikin, Wang, Wilson, and Rose]{Johnson:11}
Johnson,~M.~W.; Amin,~M. H.~S.; Gildert,~S.; Lanting,~T.; Hamze,~F.;
  Dickson,~N.; Harris,~R.; Berkley,~A.~J.; Johansson,~J.; Bunyk,~P.
  \latin{et~al.}  Quantum annealing with manufactured spins. \emph{Nature}
  \textbf{2011}, \emph{473}, 194--198\relax
\mciteBstWouldAddEndPuncttrue
\mciteSetBstMidEndSepPunct{\mcitedefaultmidpunct}
{\mcitedefaultendpunct}{\mcitedefaultseppunct}\relax
\EndOfBibitem
\bibitem[Teplukhin \latin{et~al.}(2019)Teplukhin, Kendrick, and
  Babikov]{Teplukhin:19}
Teplukhin,~A.; Kendrick,~B.~K.; Babikov,~D. Calculation of Molecular
  Vibrational Spectra on a Quantum Annealer. \emph{J. Chem. Theory Comput.}
  \textbf{2019}, \emph{15}, 4555--4563\relax
\mciteBstWouldAddEndPuncttrue
\mciteSetBstMidEndSepPunct{\mcitedefaultmidpunct}
{\mcitedefaultendpunct}{\mcitedefaultseppunct}\relax
\EndOfBibitem
\bibitem[Teplukhin \latin{et~al.}(2021)Teplukhin, Kendrick, Mniszewski, Zhang,
  Kumar, Negre, Anisimov, Tretiak, and Dub]{Teplukhin:21}
Teplukhin,~A.; Kendrick,~B.~K.; Mniszewski,~S.~M.; Zhang,~Y.; Kumar,~A.;
  Negre,~C. F.~A.; Anisimov,~P.~M.; Tretiak,~S.; Dub,~P.~A. Computing molecular
  excited states on a D-Wave quantum annealer. \emph{Sci. Rep.} \textbf{2021},
  \emph{11}, 18796\relax
\mciteBstWouldAddEndPuncttrue
\mciteSetBstMidEndSepPunct{\mcitedefaultmidpunct}
{\mcitedefaultendpunct}{\mcitedefaultseppunct}\relax
\EndOfBibitem
\bibitem[Asnaashari and Krems()Asnaashari, and Krems]{Asnaashari:23}
Asnaashari,~K.; Krems,~R. Compact quantum circuits for variational calculations
  of ro-vibrational energy levels of molecules on a quantum computer.
  \emph{arXiv:2303.09822} \relax
\mciteBstWouldAddEndPunctfalse
\mciteSetBstMidEndSepPunct{\mcitedefaultmidpunct}
{}{\mcitedefaultseppunct}\relax
\EndOfBibitem
\bibitem[Althorpe and Clary(2003)Althorpe, and Clary]{Althorpe:03}
Althorpe,~S.~C.; Clary,~D.~C. Quantum Scattering Calculations on Chemical
  Reactions. \emph{Annu. Rev. Phys. Chem.} \textbf{2003}, \emph{54},
  493--529\relax
\mciteBstWouldAddEndPuncttrue
\mciteSetBstMidEndSepPunct{\mcitedefaultmidpunct}
{\mcitedefaultendpunct}{\mcitedefaultseppunct}\relax
\EndOfBibitem
\bibitem[Clary(2008)]{Clary:08}
Clary,~D.~C. Quantum Dynamics of Chemical Reactions. \emph{Science}
  \textbf{2008}, \emph{321}, 789\relax
\mciteBstWouldAddEndPuncttrue
\mciteSetBstMidEndSepPunct{\mcitedefaultmidpunct}
{\mcitedefaultendpunct}{\mcitedefaultseppunct}\relax
\EndOfBibitem
\bibitem[Zhang and Guo(2016)Zhang, and Guo]{Zhang:16}
Zhang,~D.~H.; Guo,~H. Recent Advances in Quantum Dynamics of Bimolecular
  Reactions. \emph{Annu. Rev. Phys. Chem.} \textbf{2016}, \emph{67},
  135--158\relax
\mciteBstWouldAddEndPuncttrue
\mciteSetBstMidEndSepPunct{\mcitedefaultmidpunct}
{\mcitedefaultendpunct}{\mcitedefaultseppunct}\relax
\EndOfBibitem
\bibitem[Kaiser and Mebel(2002)Kaiser, and Mebel]{Kaiser:02}
Kaiser,~R.~I.; Mebel,~A.~M. The reactivity of ground-state carbon atoms with
  unsaturated hydrocarbons in combustion flames and in the interstellar medium.
  \emph{Int. Rev. Phys. Chem.} \textbf{2002}, \emph{21}, 307--356\relax
\mciteBstWouldAddEndPuncttrue
\mciteSetBstMidEndSepPunct{\mcitedefaultmidpunct}
{\mcitedefaultendpunct}{\mcitedefaultseppunct}\relax
\EndOfBibitem
\bibitem[Herbst and Yates(2013)Herbst, and Yates]{Herbst:13}
Herbst,~E.; Yates,~J.~T. Introduction: Astrochemistry. \emph{Chem. Rev.}
  \textbf{2013}, \emph{113}, 8707--8709\relax
\mciteBstWouldAddEndPuncttrue
\mciteSetBstMidEndSepPunct{\mcitedefaultmidpunct}
{\mcitedefaultendpunct}{\mcitedefaultseppunct}\relax
\EndOfBibitem
\bibitem[Jasper \latin{et~al.}(2014)Jasper, Pelzer, Miller, Kamarchik, Harding,
  and Klippenstein]{Jasper:14}
Jasper,~A.~W.; Pelzer,~K.~M.; Miller,~J.~A.; Kamarchik,~E.; Harding,~L.~B.;
  Klippenstein,~S.~J. Predictive a priori pressure-dependent kinetics.
  \emph{Science} \textbf{2014}, \emph{346}, 1212--1215\relax
\mciteBstWouldAddEndPuncttrue
\mciteSetBstMidEndSepPunct{\mcitedefaultmidpunct}
{\mcitedefaultendpunct}{\mcitedefaultseppunct}\relax
\EndOfBibitem
\bibitem[Klippenstein(2017)]{Klippenstein:17}
Klippenstein,~S.~J. From theoretical reaction dynamics to chemical modeling of
  combustion. \emph{Proc. Combust. Inst.} \textbf{2017}, \emph{36},
  77--111\relax
\mciteBstWouldAddEndPuncttrue
\mciteSetBstMidEndSepPunct{\mcitedefaultmidpunct}
{\mcitedefaultendpunct}{\mcitedefaultseppunct}\relax
\EndOfBibitem
\bibitem[Flynn \latin{et~al.}(1996)Flynn, Parmenter, and Wodtke]{Flynn:96}
Flynn,~G.~W.; Parmenter,~C.~S.; Wodtke,~A.~M. Vibrational Energy Transfer.
  \emph{J. Phys. Chem.} \textbf{1996}, \emph{100}, 12817--12838\relax
\mciteBstWouldAddEndPuncttrue
\mciteSetBstMidEndSepPunct{\mcitedefaultmidpunct}
{\mcitedefaultendpunct}{\mcitedefaultseppunct}\relax
\EndOfBibitem
\bibitem[Liu(2016)]{Liu:16}
Liu,~K. Vibrational Control of Bimolecular Reactions with Methane by Mode,
  Bond, and Stereo Selectivity. \emph{Annu. Rev. Phys. Chem.} \textbf{2016},
  \emph{67}, 91--111\relax
\mciteBstWouldAddEndPuncttrue
\mciteSetBstMidEndSepPunct{\mcitedefaultmidpunct}
{\mcitedefaultendpunct}{\mcitedefaultseppunct}\relax
\EndOfBibitem
\bibitem[Yang \latin{et~al.}(2019)Yang, Huang, Hu, Guo, and Xie]{Yang:19c}
Yang,~D.; Huang,~J.; Hu,~X.; Guo,~H.; Xie,~D. Breakdown of energy transfer gap
  laws revealed by full-dimensional quantum scattering between HF molecules.
  \emph{Nat. Commun.} \textbf{2019}, \emph{10}, 4658\relax
\mciteBstWouldAddEndPuncttrue
\mciteSetBstMidEndSepPunct{\mcitedefaultmidpunct}
{\mcitedefaultendpunct}{\mcitedefaultseppunct}\relax
\EndOfBibitem
\bibitem[Kim \latin{et~al.}(2015)Kim, Weichman, Sjolander, Neumark, K{\l}os,
  Alexander, and Manolopoulos]{Kim:15}
Kim,~J.~B.; Weichman,~M.~L.; Sjolander,~T.~F.; Neumark,~D.~M.; K{\l}os,~J.;
  Alexander,~M.~H.; Manolopoulos,~D.~E. Spectroscopic observation of resonances
  in the F~+~H$_2$ $\to$ HF~+~H reaction. \emph{Science} \textbf{2015},
  \emph{349}, 510\relax
\mciteBstWouldAddEndPuncttrue
\mciteSetBstMidEndSepPunct{\mcitedefaultmidpunct}
{\mcitedefaultendpunct}{\mcitedefaultseppunct}\relax
\EndOfBibitem
\bibitem[Wang \latin{et~al.}(2013)Wang, Chen, Yang, Xiao, Sun, Huang, Dai,
  Yang, and Zhang]{Wang:13b}
Wang,~T.; Chen,~J.; Yang,~T.; Xiao,~C.; Sun,~Z.; Huang,~L.; Dai,~D.; Yang,~X.;
  Zhang,~D.~H. Dynamical Resonances Accessible Only by Reagent Vibrational
  Excitation in the F~+~HD $\to$ HF~+~D Reaction. \emph{Science} \textbf{2013},
  \emph{342}, 1499\relax
\mciteBstWouldAddEndPuncttrue
\mciteSetBstMidEndSepPunct{\mcitedefaultmidpunct}
{\mcitedefaultendpunct}{\mcitedefaultseppunct}\relax
\EndOfBibitem
\bibitem[Perreault \latin{et~al.}(2017)Perreault, Mukherjee, and
  Zare]{Perreault:17}
Perreault,~W.~E.; Mukherjee,~N.; Zare,~R.~N. Quantum control of molecular
  collisions at 1 Kelvin. \emph{Science} \textbf{2017}, \emph{358}, 356\relax
\mciteBstWouldAddEndPuncttrue
\mciteSetBstMidEndSepPunct{\mcitedefaultmidpunct}
{\mcitedefaultendpunct}{\mcitedefaultseppunct}\relax
\EndOfBibitem
\bibitem[Vogels \latin{et~al.}(2015)Vogels, Onvlee, Chefdeville, van~der
  Avoird, Groenenboom, and van~de Meerakker]{Vogels:15}
Vogels,~S.~N.; Onvlee,~J.; Chefdeville,~S.; van~der Avoird,~A.;
  Groenenboom,~G.~C.; van~de Meerakker,~S. Y.~T. Imaging resonances in
  low-energy NO-He inelastic collisions. \emph{Science} \textbf{2015},
  \emph{350}, 787\relax
\mciteBstWouldAddEndPuncttrue
\mciteSetBstMidEndSepPunct{\mcitedefaultmidpunct}
{\mcitedefaultendpunct}{\mcitedefaultseppunct}\relax
\EndOfBibitem
\bibitem[Krems(2008)]{Krems:08}
Krems,~R.~V. Cold Controlled Chemistry. \emph{Phys. Chem. Chem. Phys.}
  \textbf{2008}, \emph{10}, 4079--4092\relax
\mciteBstWouldAddEndPuncttrue
\mciteSetBstMidEndSepPunct{\mcitedefaultmidpunct}
{\mcitedefaultendpunct}{\mcitedefaultseppunct}\relax
\EndOfBibitem
\bibitem[Carr \latin{et~al.}(2009)Carr, DeMille, Krems, and Ye]{Carr:09}
Carr,~L.~D.; DeMille,~D.; Krems,~R.~V.; Ye,~J. Cold and ultracold molecules:
  science, technology and applications. \emph{New J. Phys} \textbf{2009},
  \emph{11}, 055049\relax
\mciteBstWouldAddEndPuncttrue
\mciteSetBstMidEndSepPunct{\mcitedefaultmidpunct}
{\mcitedefaultendpunct}{\mcitedefaultseppunct}\relax
\EndOfBibitem
\bibitem[Balakrishnan(2016)]{Balakrishnan:16}
Balakrishnan,~N. Perspective: Ultracold molecules and the dawn of cold
  controlled chemistry. \emph{J. Chem. Phys.} \textbf{2016}, \emph{145},
  150901\relax
\mciteBstWouldAddEndPuncttrue
\mciteSetBstMidEndSepPunct{\mcitedefaultmidpunct}
{\mcitedefaultendpunct}{\mcitedefaultseppunct}\relax
\EndOfBibitem
\bibitem[Bohn \latin{et~al.}(2017)Bohn, Rey, and Ye]{Bohn:17}
Bohn,~J.~L.; Rey,~A.~M.; Ye,~J. Cold molecules: Progress in quantum engineering
  of chemistry and quantum matter. \emph{Science} \textbf{2017}, \emph{357},
  1002--1010\relax
\mciteBstWouldAddEndPuncttrue
\mciteSetBstMidEndSepPunct{\mcitedefaultmidpunct}
{\mcitedefaultendpunct}{\mcitedefaultseppunct}\relax
\EndOfBibitem
\bibitem[Devolder \latin{et~al.}(2021)Devolder, Brumer, and
  Tscherbul]{Devolder:21}
Devolder,~A.; Brumer,~P.; Tscherbul,~T.~V. Complete Quantum Coherent Control of
  Ultracold Molecular Collisions. \emph{Phys. Rev. Lett.} \textbf{2021},
  \emph{126}, 153403\relax
\mciteBstWouldAddEndPuncttrue
\mciteSetBstMidEndSepPunct{\mcitedefaultmidpunct}
{\mcitedefaultendpunct}{\mcitedefaultseppunct}\relax
\EndOfBibitem
\bibitem[Devolder \latin{et~al.}(2023)Devolder, Tscherbul, and
  Brumer]{Devolder:23}
Devolder,~A.; Tscherbul,~T.~V.; Brumer,~P. Coherent Control of Ultracold
  Molecular Collisions: The Role of Resonances. \emph{J. Phys. Chem. Lett.}
  \textbf{2023}, \emph{14}, 2171--2177\relax
\mciteBstWouldAddEndPuncttrue
\mciteSetBstMidEndSepPunct{\mcitedefaultmidpunct}
{\mcitedefaultendpunct}{\mcitedefaultseppunct}\relax
\EndOfBibitem
\bibitem[Nyman and Yu(2000)Nyman, and Yu]{Nyman:00}
Nyman,~G.; Yu,~H.-G. Quantum theory of bimolecular chemical reactions.
  \emph{Rep. Prog. Phys.} \textbf{2000}, \emph{63}, 1001\relax
\mciteBstWouldAddEndPuncttrue
\mciteSetBstMidEndSepPunct{\mcitedefaultmidpunct}
{\mcitedefaultendpunct}{\mcitedefaultseppunct}\relax
\EndOfBibitem
\bibitem[Kassal \latin{et~al.}(2008)Kassal, Jordan, Love, Mohseni, and
  Aspuru-Guzik]{Kassal:08}
Kassal,~I.; Jordan,~S.~P.; Love,~P.~J.; Mohseni,~M.; Aspuru-Guzik,~A.
  Polynomial-time quantum algorithm for the simulation of chemical dynamics.
  \emph{Proc. Natl. Acad. Sci. USA} \textbf{2008}, \emph{105},
  18681--18686\relax
\mciteBstWouldAddEndPuncttrue
\mciteSetBstMidEndSepPunct{\mcitedefaultmidpunct}
{\mcitedefaultendpunct}{\mcitedefaultseppunct}\relax
\EndOfBibitem
\bibitem[Kassal \latin{et~al.}(2011)Kassal, Whitfield, Perdomo-Ortiz, Yung, and
  Aspuru-Guzik]{Kassal:11}
Kassal,~I.; Whitfield,~J.~D.; Perdomo-Ortiz,~A.; Yung,~M.-H.; Aspuru-Guzik,~A.
  Simulating Chemistry Using Quantum Computers. \emph{Annu. Rev. Phys. Chem.}
  \textbf{2011}, \emph{62}, 185--207\relax
\mciteBstWouldAddEndPuncttrue
\mciteSetBstMidEndSepPunct{\mcitedefaultmidpunct}
{\mcitedefaultendpunct}{\mcitedefaultseppunct}\relax
\EndOfBibitem
\bibitem[Bian and Kais(2021)Bian, and Kais]{Bian:21}
Bian,~T.; Kais,~S. Quantum computing for atomic and molecular resonances.
  \emph{J. Chem. Phys.} \textbf{2021}, \emph{154}, 194107\relax
\mciteBstWouldAddEndPuncttrue
\mciteSetBstMidEndSepPunct{\mcitedefaultmidpunct}
{\mcitedefaultendpunct}{\mcitedefaultseppunct}\relax
\EndOfBibitem
\bibitem[Roggero and Carlson(2019)Roggero, and Carlson]{Roggero:2019}
Roggero,~A.; Carlson,~J. Dynamic linear response quantum algorithm. \emph{Phys.
  Rev. C} \textbf{2019}, \emph{100}, 034610\relax
\mciteBstWouldAddEndPuncttrue
\mciteSetBstMidEndSepPunct{\mcitedefaultmidpunct}
{\mcitedefaultendpunct}{\mcitedefaultseppunct}\relax
\EndOfBibitem
\bibitem[Du \latin{et~al.}(2021)Du, Vary, Zhao, and Zuo]{Du:2021}
Du,~W.; Vary,~J.~P.; Zhao,~X.; Zuo,~W. Quantum simulation of nuclear inelastic
  scattering. \emph{Phys. Rev. A} \textbf{2021}, \emph{104}, 012611\relax
\mciteBstWouldAddEndPuncttrue
\mciteSetBstMidEndSepPunct{\mcitedefaultmidpunct}
{\mcitedefaultendpunct}{\mcitedefaultseppunct}\relax
\EndOfBibitem
\bibitem[Lee \latin{et~al.}(2022)Lee, Hsieh, Zhang, and Shi]{Lee:22}
Lee,~C.-K.; Hsieh,~C.-Y.; Zhang,~S.; Shi,~L. Variational Quantum Simulation of
  Chemical Dynamics with Quantum Computers. \emph{J. Chem. Theory Comput.}
  \textbf{2022}, \emph{18}, 2105--2113\relax
\mciteBstWouldAddEndPuncttrue
\mciteSetBstMidEndSepPunct{\mcitedefaultmidpunct}
{\mcitedefaultendpunct}{\mcitedefaultseppunct}\relax
\EndOfBibitem
\bibitem[Zhang and Miller(1988)Zhang, and Miller]{Zhang:88}
Zhang,~J. Z.~H.; Miller,~W.~H. Quantum reactive scattering via the S-matrix
  version of the Kohn variational principle: Integral cross sections for
  H~+~H$_2$($v_1=j_1=0$) $\to$ H$_2$($v_2=1,\ j_2=1,3$) + H in the energy range
  $E_\text{total} = 0.9-1.4$ eV. \emph{Chem. Phys. Lett.} \textbf{1988},
  \emph{153}, 465--470\relax
\mciteBstWouldAddEndPuncttrue
\mciteSetBstMidEndSepPunct{\mcitedefaultmidpunct}
{\mcitedefaultendpunct}{\mcitedefaultseppunct}\relax
\EndOfBibitem
\bibitem[Zhang and Miller(1989)Zhang, and Miller]{Zhang:89}
Zhang,~J. Z.~H.; Miller,~W.~H. Quantum reactive scattering via the S-matrix
  version of the Kohn variational principle: Differential and integral cross
  sections for D~+~H$_2$ $\to$ HD~+~H. \emph{J. Chem. Phys.} \textbf{1989},
  \emph{91}, 1528--1547\relax
\mciteBstWouldAddEndPuncttrue
\mciteSetBstMidEndSepPunct{\mcitedefaultmidpunct}
{\mcitedefaultendpunct}{\mcitedefaultseppunct}\relax
\EndOfBibitem
\bibitem[Bravo-Prieto \latin{et~al.}()Bravo-Prieto, LaRose, Cerezo, Subasi,
  Cincio, and Coles]{VQLS}
Bravo-Prieto,~C.; LaRose,~R.; Cerezo,~M.; Subasi,~Y.; Cincio,~L.; Coles,~P.~J.
  Variational Quantum Linear Solver. \emph{arXiv:1909.05820} \relax
\mciteBstWouldAddEndPunctfalse
\mciteSetBstMidEndSepPunct{\mcitedefaultmidpunct}
{}{\mcitedefaultseppunct}\relax
\EndOfBibitem
\bibitem[Zhang \latin{et~al.}(1988)Zhang, Chu, and Miller]{Zhang:1988vy}
Zhang,~J. Z.~H.; Chu,~S.; Miller,~W.~H. Quantum scattering via the S‐matrix
  version of the Kohn variational principle. \emph{J. Chem. Phys.}
  \textbf{1988}, \emph{88}, 6233--6239\relax
\mciteBstWouldAddEndPuncttrue
\mciteSetBstMidEndSepPunct{\mcitedefaultmidpunct}
{\mcitedefaultendpunct}{\mcitedefaultseppunct}\relax
\EndOfBibitem
\bibitem[SI()]{SI}
See the supporting information associated with this article for details of
  convergence tests, additional information on the implementation of the VQLS,
  and for the basis set orthogonalization procedure.\relax
\mciteBstWouldAddEndPunctfalse
\mciteSetBstMidEndSepPunct{\mcitedefaultmidpunct}
{}{\mcitedefaultseppunct}\relax
\EndOfBibitem
\bibitem[Johnson(1973)]{Johnson:73}
Johnson,~B.~R. The multichannel Lod-Derivative Method for Scattering
  Calculations. \emph{J. Comput. Phys.} \textbf{1973}, \emph{13}, 445\relax
\mciteBstWouldAddEndPuncttrue
\mciteSetBstMidEndSepPunct{\mcitedefaultmidpunct}
{\mcitedefaultendpunct}{\mcitedefaultseppunct}\relax
\EndOfBibitem
\bibitem[Secrest and Johnson(1966)Secrest, and Johnson]{SJ-model}
Secrest,~D.; Johnson,~B.~R. Exact Quantum‐Mechanical Calculation of a
  Collinear Collision of a Particle with a Harmonic Oscillator. \emph{J. Chem.
  Phys.} \textbf{1966}, \emph{45}, 4556--4570\relax
\mciteBstWouldAddEndPuncttrue
\mciteSetBstMidEndSepPunct{\mcitedefaultmidpunct}
{\mcitedefaultendpunct}{\mcitedefaultseppunct}\relax
\EndOfBibitem
\bibitem[Stechel \latin{et~al.}(1978)Stechel, Walker, and Light]{Stechel:1978}
Stechel,~E.~B.; Walker,~R.~B.; Light,~J.~C. R‐matrix solution of coupled
  equations for inelastic scattering. \emph{J. Chem. Phys.} \textbf{1978},
  \emph{69}, 3518--3531\relax
\mciteBstWouldAddEndPuncttrue
\mciteSetBstMidEndSepPunct{\mcitedefaultmidpunct}
{\mcitedefaultendpunct}{\mcitedefaultseppunct}\relax
\EndOfBibitem
\bibitem[Manolopoulos and Gray(1995)Manolopoulos, and Gray]{Manolopoulos:95}
Manolopoulos,~D.~E.; Gray,~S.~K. Symplectic integrators for the multichannel
  Schr{\"o}dinger equation. \emph{J. Chem. Phys.} \textbf{1995}, \emph{102},
  9214--9227\relax
\mciteBstWouldAddEndPuncttrue
\mciteSetBstMidEndSepPunct{\mcitedefaultmidpunct}
{\mcitedefaultendpunct}{\mcitedefaultseppunct}\relax
\EndOfBibitem
\bibitem[Ruf and Miller(1988)Ruf, and Miller]{Ruf:88}
Ruf,~B.~A.; Miller,~W.~H. A new (Cartesian) reaction-path model for dynamics in
  polyatomic systems, with application to H-atom transfer in malonaldehyde.
  \emph{J. Chem. Soc., Faraday Trans. 2} \textbf{1988}, \emph{84},
  1523--1534\relax
\mciteBstWouldAddEndPuncttrue
\mciteSetBstMidEndSepPunct{\mcitedefaultmidpunct}
{\mcitedefaultendpunct}{\mcitedefaultseppunct}\relax
\EndOfBibitem
\bibitem[Petkovi{\'c} and K{\"u}hn(2003)Petkovi{\'c}, and
  K{\"u}hn]{Petkovic:03}
Petkovi{\'c},~M.; K{\"u}hn,~O. Multidimensional Hydrogen Bond Dynamics in
  Salicylaldimine: Coherent Nuclear Wave Packet Motion versus Intramolecular
  Vibrational Energy Redistribution. \emph{J. Phys. Chem. A} \textbf{2003},
  \emph{107}, 8458--8466\relax
\mciteBstWouldAddEndPuncttrue
\mciteSetBstMidEndSepPunct{\mcitedefaultmidpunct}
{\mcitedefaultendpunct}{\mcitedefaultseppunct}\relax
\EndOfBibitem
\bibitem[Giese and K{\"u}hn(2005)Giese, and K{\"u}hn]{Giese:05}
Giese,~K.; K{\"u}hn,~O. The all-Cartesian reaction plane Hamiltonian:
  Formulation and application to the H-atom transfer in tropolone. \emph{J.
  Chem. Phys.} \textbf{2005}, \emph{123}, 054315\relax
\mciteBstWouldAddEndPuncttrue
\mciteSetBstMidEndSepPunct{\mcitedefaultmidpunct}
{\mcitedefaultendpunct}{\mcitedefaultseppunct}\relax
\EndOfBibitem
\bibitem[Giese \latin{et~al.}(2005)Giese, Ushiyama, Takatsuka, and
  K{\"u}hn]{Giese:05b}
Giese,~K.; Ushiyama,~H.; Takatsuka,~K.; K{\"u}hn,~O. Dynamical hydrogen atom
  tunneling in dichlorotropolone: A combined quantum, semiclassical, and
  classical study. \emph{J. Chem. Phys.} \textbf{2005}, \emph{122},
  124307\relax
\mciteBstWouldAddEndPuncttrue
\mciteSetBstMidEndSepPunct{\mcitedefaultmidpunct}
{\mcitedefaultendpunct}{\mcitedefaultseppunct}\relax
\EndOfBibitem
\bibitem[Miller \latin{et~al.}(1980)Miller, Handy, and Adams]{Miller:80}
Miller,~W.~H.; Handy,~N.~C.; Adams,~J.~E. Reaction path Hamiltonian for
  polyatomic molecules. \emph{J. Chem. Phys.} \textbf{1980}, \emph{72},
  99--112\relax
\mciteBstWouldAddEndPuncttrue
\mciteSetBstMidEndSepPunct{\mcitedefaultmidpunct}
{\mcitedefaultendpunct}{\mcitedefaultseppunct}\relax
\EndOfBibitem
\bibitem[Patterson \latin{et~al.}(2010)Patterson, Tsikata, and
  Doyle]{Patterson:10}
Patterson,~D.; Tsikata,~E.; Doyle,~J.~M. Cooling and collisions of large gas
  phase molecules. \emph{Phys. Chem. Chem. Phys.} \textbf{2010}, 9736\relax
\mciteBstWouldAddEndPuncttrue
\mciteSetBstMidEndSepPunct{\mcitedefaultmidpunct}
{\mcitedefaultendpunct}{\mcitedefaultseppunct}\relax
\EndOfBibitem
\bibitem[Li and Heller(2012)Li, and Heller]{Li:12}
Li,~Z.; Heller,~E.~J. Cold collisions of complex polyatomic molecules. \emph{J.
  Chem. Phys.} \textbf{2012}, \emph{136}, 054306\relax
\mciteBstWouldAddEndPuncttrue
\mciteSetBstMidEndSepPunct{\mcitedefaultmidpunct}
{\mcitedefaultendpunct}{\mcitedefaultseppunct}\relax
\EndOfBibitem
\bibitem[Piskorski \latin{et~al.}(2014)Piskorski, Patterson, Eibenberger, and
  Doyle]{Piskorski:14}
Piskorski,~J.; Patterson,~D.; Eibenberger,~S.; Doyle,~J.~M. Cooling,
  Spectroscopy and Non-Sticking of trans-Stilbene and Nile Red.
  \emph{ChemPhysChem} \textbf{2014}, \emph{15}, 3800--3804\relax
\mciteBstWouldAddEndPuncttrue
\mciteSetBstMidEndSepPunct{\mcitedefaultmidpunct}
{\mcitedefaultendpunct}{\mcitedefaultseppunct}\relax
\EndOfBibitem
\end{mcitethebibliography}

\end{document}


\singlespacing

\maketitle

In this Supporting Information (SI), we present technical details regarding basis set convergence of $S$-matrix elements  (Sec. S1), and the implementation of the VQLS  algorithm for a model $2\times 2$ matrix (Sec. S2). Section S3 presents an overview of the basis set orthogonalization procedure used in our Q-KVP calculations  for the SJ model.  

\newpage 
\section{Convergence tests}

 \begin{figure}[t]
	\centering
	\includegraphics[width=12 cm]{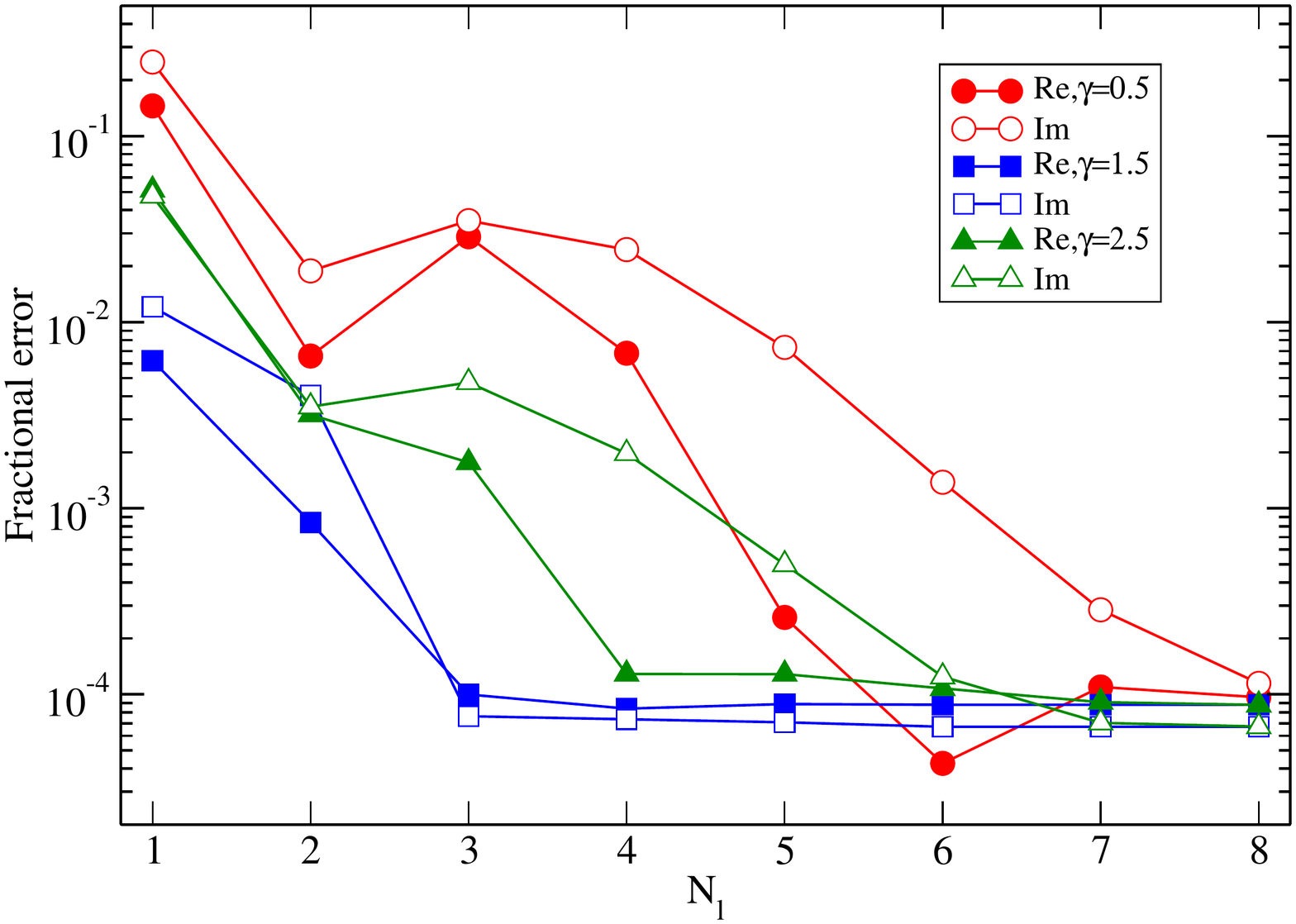}
	\caption{Convergence of the real (Re)  and  imaginary (Im) parts of  the S matrix  as a function of the number of $L^2$ basis functions for the single-channel scattering problem defined in the main text. }
	\label{fig:1D}
\end{figure}

Figure~\ref{fig:1D} shows the convergence of  the $S$-matrix elements for the one-dimensional scattering problem defined in the main text.  The real and imaginary parts of the $S$-matrix are plotted as a function of $N_l$, the number of $L^2$ basis functions. The fractional error  $ (S_\text{CC}-S_\text{Q-KVP})/S_\text{CC}$ is defined with respect to  the $S$-matrix elements obtained from exact CC calculations. The fractional error decreases with increasing $N_l$ saturating at about $10^{-4}$ for all basis set exponents $\gamma$ (see Eq. (3) of the main text).  We also note that certain choices of the $\gamma$ parameter can lead to faster basis set convergence. Notably, the fractional error is less than 0.02 even for the smallest basis set, which includes two real basis functions, which implies that the one-channel scattering problem can be solved by inverting a $2\times 2$ $\bf M$ matrix.

\section{VQLS matrix inversion}

We invert the $2\times 2$ $\bf M$ matrix  on the Qiskit platform using the VQLS algorithm. The optimal depth of the ansatz is estimated based on the fidelity and computing time for $E=0.55$ in Table \ref{tab:S1}. We observe that a very good fidelity of 0.999 can be obtained  for a  $2\times 2$ $\bf M$ matrix using only a single-layer ansatz, with the computational cost $\tau$ of less than a second.  

 \begin{center}
\begin{table}[t]
\renewcommand{\arraystretch}{1.2} \addtolength{\tabcolsep}{5 pt}
\begin{tabular}{ccccccc}
\hline \hline
$k$ &$\mathcal{F}^1$ &$\mathcal{F}^2$&$\mathcal{F}^3$& $\tau^1$ (s) &$\tau^2$ (s) &$\tau^3$ (s) \\
\hline 
1 & 0.999 &0.999&0.999&0.162&0.339&0.441\\
2 & 0.999 &0.999&0.999&0.203&0.441&0.644\\
\hline \hline
\end{tabular}
\caption{The  fidelity $\mathcal{F}$ and  computing  time $\tau$ of the VQLS computation of the $k$-th column of the $2\times 2$ $\bf M$ matrix at $E=0.55$ for the single-channel scattering problem  defined in the main text. The superscripts $d$ of $\mathcal{F}^{(d)}$ and  $\tau^{(d)}$  indicates the number of layers in the ansatz.  }
\label{tab:S1}
\end{table}
\end{center}

\section{Orthogonalized finite basis set representation }

Because the $L^2$ basis functions used in the S-matrix version of the KVP  are  generally not orthogonal, an overcompleteness problem can arise. To overcome this problem, we orthogonalize the  basis using a symmetric orthogonalization technique. Specifically, for the 2D SJ model (see the main text) we select the eigenvalues $\lambda_i$ of the overlap matrix
 \begin{equation}\label{overlap}
O_{ln,l'n'}= \langle u_{ln}\phi_{n}|u_{l'n'}\phi_{n'} \rangle
 \end{equation}
 greater than a predetermined cutoff value to build $N_q \le N(N_l-1)$ orthogonalized  basis functions
  \begin{equation}\label{orthogonalized_fs}
\ket{\Theta_{i}}=\dfrac{1}{\sqrt{\lambda_i}}\sum_{l,n}\bar{X}_{ln,i}|u_{ln}\phi_n\rangle,
 \end{equation}
 where $\bar{X}_{ln,i}$ are the eigenvector components of the overlap matrix \eqref{overlap}. 
 The number of orthogonalized basis functions depends on  the cutoff parameter and the number of primitive basis functions.
A rectangular transformation matrix $\bf{X}$ with elements ${X}_{ln,i}={\bar{X}_{ln,i}}/{\sqrt{\lambda_i}} \,\, (i =1,\ldots, N_q)$ is then used to transform  the matrices $\bf M$ and $\mathbf{M}_0$  (see main text)  to the orthogonalized basis as $\bf \tilde{M}=X^TMX$ and $\bf \tilde{M}_0= X^TM_0$, respectively. The transformed matrices, which do not suffer from the overcompleteness problem, have dimensions $N_q  \times N_q $ for $\bf \Tilde{M}$ and $ N_q   \times N_o$ for $\bf \tilde{M}_0$, where $N_o$ is the number of open channels.